\documentclass{elsarticle}

\usepackage{lineno,hyperref}

\usepackage{xr}

\journal{Journal of \LaTeX\ Templates}

\newcommand{\be}{\begin{equation}}
\newcommand{\ee}{\end{equation}}
\newcommand{\la}{\langle}
\newcommand{\ra}{\rangle}

\newcommand{\e}{{\rm{e}}}

\newcommand{\lf}{\left}
\newcommand{\rg}{\right}

\newcommand{\Lat}{L_{\alpha}^{\theta}}

\usepackage{bbm}
\usepackage{color}
\usepackage{bigints}
\usepackage{graphicx}
\usepackage{subfig}

\begin{document}

\begin{frontmatter}

  \title{Langevin equation in complex media \\
    and anomalous diffusion
}
%
%


\author[UNIBO]{Silvia Vitali}
\author[POTSDAM,BCAM]{Vittoria Sposini}
\author[BCAM]{Oleksii Sliusarenko}
\author[ISTI,BCAM]{Paolo Paradisi\corref{correspondingauthor1}}
\cortext[correspondingauthor1]{Corresponding author}
\ead{paolo.paradisi@cnr.it}
\author[UNIBO]{Gastone Castellani}
\author[BCAM,Ikerbasque]{Gianni Pagnini\corref{correspondingauthor2}}
\cortext[correspondingauthor2]{Corresponding author}
\ead{gpagnini@bcamath.org}


\address[UNIBO]{Department of Physics and Astronomy, Bologna University,
  Viale Berti Pichat 6/2, I-40126 Bologna, Italy}
\address[POTSDAM]{Institute for Physics and Astronomy, University of Potsdam,
  Karl-Liebknecht-Strasse 24/25, D-14476 Potsdam-Golm, Germany}
\address[BCAM]{BCAM--Basque Center for Applied Mathematics, Alameda de 
Mazarredo 14, E-48009 Bilbao, Basque Country, Spain}
\address[ISTI]{ISTI--CNR, Institute of Information Science and Technologies
  "A. Faedo” (Consiglio Nazionale delle Ricerche),
  Via Moruzzi 1, I-56124 Pisa, Italy}
\address[Ikerbasque]{Ikerbasque--Basque Foundation for Science, Calle de 
Mar\'ia D\'iaz de Haro 3, E-48013 Bilbao, Basque Country, Spain}

\begin{abstract}
The problem of biological motion is a very intriguing and topical issue.
%
%
%
Many efforts are being focused on the development of novel modeling 
approaches for the description of anomalous diffusion in biological systems, 
such as the very complex and heterogeneous cell environment.
Nevertheless, many questions are still open, such as the joint manifestation
of statistical features in agreement with different models that
can be also somewhat alternative to each other,
e.g., Continuous Time Random Walk (CTRW) and Fractional Brownian Motion (FBM).
%
%
To overcome these limitations, we propose a stochastic diffusion model
with 
additive noise and linear friction force (linear Langevin equation), thus
involving the explicit modeling of velocity dynamics.
The complexity of the medium
is parameterized via a population of intensity parameters (relaxation time 
and diffusivity of velocity), thus introducing an additional randomness, in addition
to white noise, in the particle's dynamics.
We prove that, for proper distributions of these parameters, we can get
both Gaussian anomalous diffusion, fractional diffusion and its
generalizations.
%
%
%
%
\end{abstract}


\end{frontmatter}


\section{Introduction}

\noindent
The very rich dynamics of biosystem movements have been attracting the interest 
of many researchers in the field of statistical physics and complexity
for its inherent temporal and spatial multi-scale character. 
Further, new techniques allowed to track the motion of large biomolecule in 
the cell with great temporal and spatial 
accuracy, both {\it in vivo} and {\it in vitro}
\cite{hofling_etal-rpphys-2013,regner_etal-biophysj-2013,manzo_etal-rpp-2015}.
%
Two main transport mechanisms were identified: 
(i) passive motion, determined by the cytoplasm crowding and (ii) active 
transport, given by the presence of molecular motors carrying biomolecules along
filaments and microtubules (cytoskeleton)
\cite{tolicnorrelykke_etal-prl-2004,golding_etal-prl-2006,javer_etal-natcomm-2014,caspi_etal-prl2000}.
Diffusion processes have been used to describe many biological phenomena such 
as molecular motion through cellular membrane
\cite{weigel_etal-pnas-2011,javanainen_etal-fd-2013,krapf_etal-pccp-2016,metzler_etal-bba-2016},
DNA motility within cellular nucleus \cite{javer_etal-natcomm-2014},
chromosome dynamics and motility on fractal DNA 
globules \cite{tamm_etal-prl-2015}, motion of mRNA molecules in 
{\it Escherichia Coli} bacteria \cite{golding_etal-prl-2006} and of lipid 
granules in yeast cells \cite{tolicnorrelykke_etal-prl-2004}.

Standard or normal diffusive (Brownian) motion is uniquely described by the
Wiener process \cite{risken-1989} and is associated with a Gaussian 
Probability Density Function (PDF) of displacements and linear time
dependence of the Mean Square Displacement (MSD).
It is well-known that normal diffusion emerges in the long-time limit 
$t \gg \tau_c$ when the correlation time scale $\tau_c$ is finite and
non-zero \cite{taylor-1921}
(see Section \ref{random_media} of Supplementary Material for details). 
However, biosystems' diffusion is often non-standard, with non-Gaussian
PDF of displacements and non-linear time dependence of MSD:
\be
\sigma^2_X(t) = \la (X_t-X_0)^2 \ra \sim D_\phi \, t^\phi\ \ ;\quad \phi>0\ ,
\label{anom_diff}
\ee
where $X(t)$ is the position. This is known as 
{\it anomalous diffusion}, distinguished in slow {\it subdiffusion} 
($\phi < 1$) and fast {\it superdiffusion} ($\phi > 1$).

Normal diffusion is recovered for $\phi=1$.
  
%


The general condition for anomalous diffusion to occur is to have a zero or
infinite $\tau_c$ \cite{taylor-1921} and, precisely:
\begin{itemize}
\item
{\it Superdiffusion}:
\be
\tau_c = \infty:\ \langle X^2\rangle\sim t^\phi\ \ {\rm with}\ \ 1 < \phi \le 2 
\ \ {\rm or}\ \ \langle X^2\rangle = \infty\ .
\label{superdiff1}
\ee
\item
{\it Subdiffusion}:
\be
\tau_c = 0:\ \langle X^2\rangle\sim t^\phi\ \  {\rm with}\ \  0 < \phi \le 1 \ .
\label{subdiff1}
\ee
\end{itemize}

(see Section \ref{random_media} of Supplementary Material for a detailed
discussion about this point).

\noindent
Both subdiffusion and superdiffusion have been found in cell transport, the
first 
one being usually related to passive motion and the latter one to active motion
(see, e.g., Refs. \cite{tolicnorrelykke_etal-prl-2004,golding_etal-prl-2006,caspi_etal-pre-2002,bronstein_etal-prl-2009} for subdiffusion, and Refs.
\cite{javer_etal-natcomm-2014,caspi_etal-prl2000,robert_etal-plos1-2010,reverey_etal-scirep-2015} for superdiffusion).

At variance with normal diffusion different physical/biological conditions can 
originate anomalous diffusion \cite{bouchaud_etal-physrep-1990,metzler_etal-physrep-2000} and several models and interpretations were proposed in the recent 
literature
\cite{hofling_etal-rpphys-2013,manzo_etal-rpp-2015,burov_etal-pccp-2011,metzler_etal-pccp-2014}.
Widely investigated models of anomalous diffusion are Continuous Time Random 
Walk (CTRW) \cite{metzler_etal-physrep-2000} and
Fractional Brownian Motion (FBM) \cite{mandelbrot_etal-siamrev-1968}, both
models sharing the same  anomalous diffusive scaling of Eq. (\ref{anom_diff}).
Many authors compared these models with each other and with data, essentially 
finding some features to be satisfied by the CTRW (weak ergodicity breaking and
aging) \cite{burov_etal-pccp-2011,he_etal-prl-2008,jeon_etal-prl-2011} and other
 ones by the FBM (e.g., the p-variation index 
\cite{magdziarz_etal-prl-2009,kepten_etal-pre-2011,burnecki_etal-biophj-2012}).
Despite the efforts of many research groups, an exhaustive model explaining all 
the statistical features of experimental data does not yet exist and
the research is recently focusing on alternative approaches, such as
Heterogeneous Diffusivity Processes (HDPs)
\cite{cherstvy_etal-njp-2013,massignan_etal-prl-2014,chubynsky_etal-prl-2014,cherstvy_etal-pccp-2016,sposini_njp18}
       or other  similar approaches based on
fluctuations of some dynamical parameter, e.g.,
fluctuating friction governed  by a stochastic differential equation
\cite{rozenfeld_etal-pla-1998,luczka_etal-pa-2000,luczka_etal-appb-2004},
mass of a Brownian-like particle randomly fluctuating in the course of time
\cite{ausloos_etal-pre-2006}.
%

All these approaches
can be linked to superstatistics \cite{beck-prl-2001,beck_etal-pa-2003},
%
whose main idea is that of a complex inhomogeneous environment divided into
cells,
each one characterized by a nearly uniform value of some intensive parameters.
Then, a Brownian test particle experiences parameter fluctuations during a cell-to-cell
transition \cite{beck_etal-pa-2003}. 
%
%
In general, superstatistics is successful to model:
turbulent dispersion (energy dissipation fluctuations) \cite{beck-prl-2001},
renewal critical events in intermittent systems \cite{paradisi_cejp09,akin_jsmte09}
and, for different distributions of the fluctuating intensive quantities,
different effective statistical mechanics can be derived \cite{beck_etal-pa-2003}, 
e.g., Tsallis statistics with $\chi^2$-distribution \cite{beck-prl-2001}.
Diffusing Diffusivity Models (DDMs), with position diffusivity governed by a
stochastic differential equation,
are being recently proposed \cite{chubynsky_etal-prl-2014} and are attracting the interest
of many authors as they represent an important attempt to go beyond superstatistics
\cite{sposini_njp18,jain_jcs17,chechkin_etal-prx-2017,lanoiselee_jpamt18}.

In this framework, we propose a modeling approach to anomalous diffusion inspired by
the constructive approach used to derive the Schneider grey noise, the grey Brownian Motion (gBM)
\cite{schneider_1990,schneider_1992} and the generalized grey Brownian Motion
(ggBM)
\cite{mura-phd-2008,mura_etal-pa-2008,mura_etal-jpa-2008,mura_etal-itsf-2009,pagnini_etal-ijsa-2012,pagnini_etal-ptrsa-2013}
  (see Section \ref{grey_noise} of Supplementary Material for
  a brief survey about grey noise, gBM and ggBM).
Such processes emerge to be equivalent to the product of the FBM $B_H(t)$ with
an independent positive random variable $\lambda$,
i.e., the amplitude associated to each single trajectory can change from one
trajectory to another one ($H$ is the self-similarity Hurst exponent).
%
%
%
%
When the amplitude PDF is the Mainardi distribution $M_\beta(\lambda)$ with
properly chosen scaling $\beta$ (depending on the FBM scaling $H$)
\cite{mainardi_etal-ijde-2010,pagnini-fcaa-2013,pagnini-pa-2014},
grey noise is a stochastic solution of the Time Fractional Diffusion Equation 
(TFDE) \cite{gorenflo_etal-cp-2002,gorenflo_etal-pa-2002,mainardi_etal-fcaa-2001}, i.e, the gBM-PDF $P(x,t)$ is a solution of the TFDE
(see Section \ref{mainardi} of Supplementary Material
for a brief survey about the Mainardi function).
The ggBM generalizes gBM by considering independent scaling parameters $\beta$ and $H$
and it was recently recognized to be a stochastic solution of the Erd\'elyi--Kober
Fractional Diffusion Equation (EKFDE) \cite{pagnini-fcaa-2012}.
A further extension of the ggBM is given by the process introduced in
Ref. \cite{pagnini_etal-fcaa-2016}, where the
amplitude distribution is generalized to a combination of L\'evy distributions
by imposing the ggBM-PDF to be compatible with the Space-Time Fractional
Diffusion Equation (STFDE)
\cite{gorenflo_etal-cp-2002,gorenflo_etal-pa-2002,mainardi_etal-fcaa-2001,li_jcnld17}. 
%
%
Interestingly, ggBM can also describe nonstationary and aging behaviors.
The potential applications of ggBM to biological transport were recently 
discussed in Ref. \cite{molina_etal-pre-2016}, where the ggBM compatible with
EKFDE was investigated by means of several statistical indices commonly used
in the analysis of particle tracking data. The authors showed that the ggBM
approach accounts for the weak ergodicity breaking and aging 
(CTRW) and, at the same time, for the p-variation test (FBM).
A DDM and a ggBm-like model (namely a randomly-scaled Gaussian process)
with random position diffusivity governed by the same stochastic equation have been recently
compared each other \cite{sposini_njp18}.

\noindent
However, the physical interpretation of ggBM approach based on the FBM
is not completely clear. Further, potential applications to transport
in a viscous fluid needs to include at least the effect of viscosity.

\vspace{.5cm}
\noindent
In order to include the effect of viscosity, we describe the development of a
model similar to the original ggBM, but with a 
friction-diffusion process instead of a Gaussian noise, thus involving an
explicit modeling of system's dynamics by substituting the FBM, used to built
the ggBM, with the stochastic process resulting from
Langevin equation for the particle velocity.
%
In particular,
we use a Langevin equation with a linear viscous term (Stokes drag) and an 
additive white Gaussian noise, also known as Ornstein--Uhlenbeck (OU) process 
\cite{risken-1989}. 
The system's complexity is described by proper random fluctuations 
of the parameters in the velocity Langevin equation:
relaxation time, related to friction; velocity diffusivity, related to noise
intensity.
It is worth noting that the medium is here composed of the underlying fluid
substrate and of the particle ensemble.
Medium complexity is then not mimicked by random temporal fluctuations, but
described by
inter-particle fluctuations of parameters and, thus, by proper time-independent
statistical distributions that characterize the complex medium.
%
In next sections we show that this assumption allows to get anomalous diffusion
if proper parameter distributions are chosen. 
In this sense, this model also generalizes 
the approach of HDPs as it also accounts for the heterogeneity of the friction 
parameter, thus including the effect of relaxation due to viscosity that,
in other HDPs, is completely neglected.
In this work we 
focus on superdiffusion, which is derived for a free particle motion by means 
of a general argument.
%
%

\vspace{.5cm}
The paper is organized as follows. 
In Section \ref{super_diff} we introduce the randomized Langevin model for 
superdiffusion, based on the free motion of Brownian particles in
a viscous medium.
In Section \ref{simulations} we show the results of numerical simulations.
In particular, we numerically test some crucial assumptions, such as the 
existence of a generalized equilibrium/stationary condition in the long-time 
limit.
In Section \ref{conclusion} we sketch some conclusions and discuss
the potential applications of the proposed model.
Mathematical details can be found in the Supplementary Material.

\section{Free particle motion and superdiffusion}
\label{super_diff}

\noindent
Consider the following linear Langevin equation for the
velocity $V(t)$ of a particle moving in a viscous medium:
\be
\frac{dV_t}{dt} = -\frac{V_t}{\tau} + \sqrt{2 \nu}\ \xi_t
\label{lang_free}
\ee
%
%
being $\tau$ the relaxation time scale\footnote{
Given the particle mass $m$ and the friction coefficient $\gamma$, it
results: $\tau=m/\gamma$.
},
and $\nu$ the velocity diffusivity, which has dimensional units:
$[\nu] = [V^2]/[T]$.
The diffusivity $\nu$ determines the intensity of the Gaussian 
white noise $\xi_t$.
This is a random uncorrelated force:
\be
\langle \xi_t \rangle = 0\ ;\quad 
\langle \xi_t \cdot \xi_{t^\prime} \rangle = \delta(t-t^\prime)\ ,
\label{white_noise}
\ee
whose stochastic It\^o integral is a Wiener process \cite{risken-1989}.
When $\tau$ and $\nu$ are fixed parameters, Eq. (\ref{lang_free}) is a 
OU process (see, e.g., \cite{risken-1989}),
which, together with the kinematic equation:
\be
\frac{dX_t}{dt} = V_t\ ,
\label{diff_eq1}
\ee 
is the most simple stochastic model for the one-dimensional free motion of a 
particle in a viscous medium, with thermal fluctuations depicted by the white
noise $\xi_t$.

In the Langevin model with random parameters here proposed, single path 
dynamics are given by Eq. (\ref{lang_free}), but the statistical ensemble of 
paths is affected not only by randomness in the white noise $\xi_t$, but  also
in the parameters $\tau$ and $\nu$, whose randomness describes the
complex medium.
In order to derive the overall statistical features of $X_t$ and $V_t$, 
the  computation is carried out in three steps.
First we consider the averaging operation with respect to the noise term 
$\xi_t$ and how the presence of a population for the parameters $\tau$ and $\nu$ affects some statistical properties of the process.
Then we consider the average over the random parameter $\tau$ and we evaluate the
PDF $g(\tau)$ in  order to get 
an anomalous superdiffusive scaling.
Finally we evaluate the PDF $f(\nu)$ in order to get the distribution 
$P(x,t)$ compatible with fractional diffusion, i.e., equal to
the fundamental solutions of some class of fractional diffusion equations 
\cite{pagnini-fcaa-2012,pagnini_etal-fcaa-2016}, or with other kinds of
diffusion processes.

The averaging operation with respect to the noise term 
$\xi_t$  gives the statistical features conditioned to the random
parameters $\tau$ and $\nu$, which result to be exactly the same as the 
standard OU process as shown in Box 1.
In particular, we are interested in the stationary correlation function
conditioned to $\tau$ and $\nu$, which reads
(see Eqs. (\ref{corr_v_stat}) and (\ref{var_stat}), Box 1):
\be 
R( t | V_0,\tau,\nu) =\nu\tau e^{-t/\tau} \,.
\label{stat_corr}
\ee
Given Eq. (\ref{stat_corr}) and considering statistically independent populations of $\tau$
and $\nu$, the stationary correlation function of the ensemble is given by:
\be
R(t) = \left\langle \nu \right\rangle 
\left\langle \tau e^{-t/\tau} \right\rangle = \int_0^\infty \nu  f(\nu) d\nu \cdot 
\int_0^\infty \ \tau\ e^{-t/\tau}\ g(\tau) d\tau 
\, ,
\label{superpos_corr_eq}
\ee
where $g(\tau)$ and $f(\nu)$ are the PDFs of the parameters $\tau$
and $\nu$, respectively.
The conditional MSD is derived from the conditional correlation function
$R( t | V_0,\tau,\nu)$, Eq. (\ref{stat_corr}), and, accordingly, 
the effective or global MSD (averaged over $\tau$ and $\nu$), is derived from  
the global correlation function R(t), Eq. (\ref{superpos_corr_eq})
(see Section \ref{random_media} of Supplementary Material).
The standard OU process is recovered for $f(\nu) = \delta (\nu-\overline{\nu})$
and $g(\tau) = \delta (\tau-\overline{\tau})$, that is, when the parameters
$\nu$ and $\tau$ are the same for all trajectories.

Does such stationarity correspond to an equilibrium condition ? 
An equilibrium state is defined by the equilibrium velocity distribution, 
which is independent of the initial conditions and it is reached by the 
system after a transient time.
When equilibrium is reached, the process becomes stationary: the 
nonstationary term of the correlation function becomes negligible and only the 
stationary correlation given in Eq. (\ref{stat_corr}) survives.
The decay of the nonstationary correlation term corresponds rigorously to 
equilibrium in the standard OU process with fixed $\tau$ and $\nu$ as shown in
Box 1. However, it is not straightforward that this feature 
also extends to the Langevin equation with random parameters, 
Eq. (\ref{lang_free}).

It is worth noting that the average of the {\it conditional stationary velocity variance} (Eq. (\ref{var_stat}), Box 1) over $\tau$ and $\nu$ gives:
\be
\langle V^2 \rangle_{\rm st} = R(0) = \langle \nu \rangle \langle \tau \rangle
\ ,
\label{var_stat_av}
\ee
which resembles an equilibrium condition extending that of the standard OU 
process, by considering the mean values of
$\tau$ and $\nu$.
This condition cannot be assumed {\it a priori}, but, if equilibrium exists,
it surely needs a stationary assumption, so that, in the following, we assume
that, in the long-time regime $t_1, t_2 \gg \langle\tau\rangle$, 
the stationary state defined by Eq. (\ref{var_stat_av}) is reached within a 
good approximation.
Consequently, in this model we consider an approximated stationary condition by setting to
zero the non-stationary term of the correlation function in Eq. (\ref{cond_corr_ou}) (Box 1).
The validity of the stationary assumption and its coincidence with the
emergence of an equilibrium distribution will be discussed later and verified
by means of numerical simulations\footnote{
The existence of an equilibrium distribution is actually verified by means of 
numerical simulations and it is also shown to coincide with the validity of 
Eq. (\ref{var_stat_av}) in the long-time regime. 
As a consequence, by applying the average over $\tau$ and $\nu$ to the
conditional velocity correlation function (Eq. (\ref{cond_corr_ou}), Box 1),
we find that the first term is exactly zero when the 
initial velocity distribution is the equilibrium one and this proves that
our model is self-consistent.
}
(see Subsection \ref{results}).

\vspace{.2cm}
\noindent\fbox{%
    \parbox{\textwidth}{%
    \small
        {\bf Box 1. OU statistics conditioned to $\tau$ and $\nu$}
        
\noindent
The statistical features conditioned to the values of $\tau$ and $\nu$ are 
given by the same mathematical
expressions of the standard OU process \cite{risken-1989}. 

Given the initial condition $V_0 = V(0)$, the solution for $t \ge 0$ of 
Eq. (\ref{lang_free}) is given by:
\be
V_t = e^{-t/\tau} \left[ V_0 + \sqrt{2 \nu} \int_{0}^t e^{t^\prime/\tau}
  \xi_{t^\prime} dt^\prime \right] \ .
\label{ou_formal_sol}
\ee
This solution can be exploited to derive the {\it conditional velocity
correlation function}, where the average is here made over the noise $\xi_t$:
\be
\langle V_{t_1} \cdot V_{t_2}|V_0,\tau,\nu \rangle = 
\left(V_0^2 - \nu \tau \right) e^{-(t_1+t_2)/\tau}
+ \nu \tau e^{-|t_1-t_2|/\tau}\ .
\label{cond_corr_ou}
\ee
The conditional 
dependence of the average on the initial velocity $V_0$ and on the
parameters $\tau$ and $\nu$ has been explicitly written. 
The choice of the initial velocity 
distribution affects the way the system relaxes to the equilibrium condition,
but not the equilibrium condition itself.
The correlation function includes two terms: the first one is the 
nonstationary transient associated with the memory of the initial condition 
$V_0$, while the second one is the stationary component depending only on the 
time lag between $t_1$ and $t_2$. In the long time limit $t_1,t_2\gg \tau$, the
first term becomes negligible, thus giving the 
{\it conditional stationary correlation function}:
\be
R( t | V_0,\tau,\nu) = \langle V_{t_1} \cdot V_{t_1+t}|V_0,\tau,\nu \rangle = 
R(0 | V_0,\tau,\nu) e^{-t/\tau} \ ,
\label{corr_v_stat}
\ee
being $t = |t_2-t_1|$ the time lag and:
\be
R(0 | V_0,\tau,\nu) = \langle V^2 | V_0,\tau,\nu \rangle_{\rm st} = \nu \tau\ ,
\label{var_stat}
\ee
the {\it conditional stationary velocity variance}, which results to be 
independent of time $t_1$ and of the initial velocity $V_0$.
 }
}
%
The correlation 
function $R(t)$ defined in Eq. (\ref{superpos_corr_eq}) and the PDF $g(\tau)$ 
must satisfy a list of features
to describe superdiffusion, i.e., 
$\sigma^2_X(t) \sim t^\phi;\,R(t) \sim t^{\phi-2};\, 1 < \phi < 2$, concerning the asymptotic time scaling of the functions, 
normalization and finite mean conditions for the distribution of time
scales $g(\tau)$
(see Box 2 in Supplementary Material).

It is worth noting that the statistical distribution of $\nu$ does not affect 
the scaling of the correlation function in Eq.(\ref{superpos_corr_eq}), but it
only introduces a multiplicative factor.
Therefore a constructive approach similar to that adopted to built up the 
{\it generalized grey Brownian motion}
\cite{mura-phd-2008,mura_etal-jpa-2008,mura_etal-itsf-2009,pagnini_etal-fcaa-2016} can be applied to our model, 
randomness of $\tau$ determining the anomalous diffusion scaling and that 
of $\nu$ the non-Gaussianity of both velocity and position distributions.

Regarding the PDF $g(\tau)$, the following: 
%
%
\be
g(\tau)=\frac{\eta}{\Gamma(1/\eta)}\frac{1}{\tau}L_\eta^{-\eta}
\lf( \frac{\eta}{\Gamma(1/\eta)}\frac{\tau}{\langle\tau\rangle} \rg)\ ;
\quad 0 < \eta < 1 \, ,
\label{pdf_gtau}
\ee
indeed satisfies all the required constrains (i-iv) listed in the
Supplementary Material (Box 2, proofs in Section \ref{appendix_gtau}).
We stress that the choice of $g(\tau)$ is not arbitrary, but addressed
(not derived) by the required constrains listed in Box 2 of Supplementary
Material.

\vspace{.2cm}
\noindent
In the above expression, $g(\tau)$ depends on the parameter $\eta$, which is
the index of the L\'evy stable, unilateral PDF $L_\eta^{-\eta}$, and on the mean 
relaxation time scale ${\langle\tau\rangle}$.
With the above choice, we get the following asymptotic behavior for
the stationary correlation function, conditioned to $\nu$, when 
$t\rightarrow\infty (t \gg \langle\tau\rangle)$ 
(see Section \ref{appendix_gtau} of Supplementary Material for details):
\begin{equation}
R(t | \nu ) = \nu\, \frac{\Gamma( 1 + \eta )}{\Gamma( 1 - \eta )}
\left( \frac{\Gamma(1/\eta)}{\eta} \right)^\eta
\la \tau \ra^{1+\eta} \, t^{-\eta} \ .
\label{corr_free}
\end{equation}
By applying Eq. (\ref{taylor21_1}, Supplementary Material) we get the 
(superdiffusive) scaling for the MSD: $\sigma^2_X(t|\nu) \propto t^\phi$ with $1 < \phi=2-\eta < 2$.

Notice that the calculations are here made under the assumption of the
approximated stationary condition discussed previously.
In this regime, $X(t)$ is exactly a Gaussian variable, as it can be reduced to 
a sum, over time, of almost independent Gaussian distributed velocity 
increments.
%
Eq. (\ref{taylor21_1}) (or, equivalently, Eq. (\ref{taylor21_2})) 
in Supplementary Material, which is essentially a sum of variances
of Gaussian distributed variables, so that the overall effect of
$g(\tau)$ is the emergence of a Gaussian variable with the anomalous, 
nonlinear, scaling of the variance given in Eq. (\ref{superdiff1})\footnote{
It is worth noting that the random superposition of Langevin equations with
randomized $\tau$ is an example of a Gaussian process with anomalous diffusion 
scaling that is different from the standard fractional Brownian motion (fBm).
}.
The resulting PDF of $X_t$ conditioned to $\nu$ is then given by the 
following Gaussian law:
\begin{eqnarray}
P(x,t|\nu)={\cal G}(x,\sigma_X^2(t | \nu)) &=& \frac{1}{\sqrt{2 \pi \sigma_X^2(t | \nu)}}
\exp\left\{ -\frac{x^2}{2\sigma_X^2(t | \nu)} \right\}\ ;
\label{gauss_super_1} 
\\
\ &&  \nonumber \\
\sigma_X^2(t | \nu) &=& 2\, C\, \nu\, t^\phi\ ; 
\quad 1 < \phi = 2-\eta < 2\ ;
\label{gauss_super_2} \\
\ && \nonumber \\
C  &=& 
\frac{\Gamma(\eta+1)}{\Gamma(3-\eta )} 
\lf( \frac{\Gamma(1/\eta)}{\eta}\rg)^{\eta}\la\tau\ra^{1+\eta}\ . 
\label{gauss_super_3}
\end{eqnarray}

The conditional dependence of ${\cal G}(x,t|\nu)$ on $\nu$ is clearly 
included in $\sigma_X^2(t | \nu)$.
The one-time PDF of the diffusion variable $X_t$ is given by the 
application of the conditional probability formula:
\be
P(x,t) = \int_0^\infty {\cal G}(x,2 C \nu t^\phi) f(\nu) d\nu = 
\int_0^\infty \frac{ \exp \left\{ -\frac{x^2}{4Ct^\phi\nu} \right\} }
{ \sqrt{4 \pi C t^\phi \nu} } f(\nu) d\nu\ .
\label{pdf_nu_super}
\ee
This relationship is formally similar to Eq. (3.9) of Ref. 
\cite{pagnini_etal-fcaa-2016}. Thus, comparing with this same equation and
after some algebraic manipulation,
Eq. (\ref{pdf_nu_super}) can be generalized to the following general form by
including the scaling exponent $\phi$:
\be
\frac{1}{ \left( C\, \overline{\nu}\, t^\phi \right)^{1/2} } 
K_{\alpha,\beta}^{0} \left( \frac{x}{\left( C \overline{\nu} t^\phi \right)^{1/2} }
\right)
 = \int_0^\infty {\cal G}(x,2\, C\, \nu\, t^\phi) \frac{1}{\overline{\nu}} 
K_{\alpha/2,\beta}^{-\alpha/2}\left( \frac{\nu}{\overline{\nu}} \right) d\nu \ ,
\label{random_nu}
\ee
with $1 < \phi=2-\eta < 2$, $f(\nu) = \frac{1}{\overline{\nu}} 
K_{_{\alpha/2,\beta}}^{_{-\alpha/2}}(\frac{\nu}{\overline{\nu}})$ and 
$C=C(\eta,\langle \tau \rangle)$ given by Eq. (\ref{gauss_super_3}).
The reference scale $\overline{\nu}$ is needed to give the proper 
physical dimensions to the random velocity diffusivity $\nu$.
As we consider only symmetric diffusion, $\theta=0$, the general range of 
parameters $\alpha$ and $\beta$ is given by:
\be 
\hspace{-0.1truecm}
0<\alpha\le 2 \,, \quad
%
%
0 < \beta \le 1 \quad
\hbox{or} \quad 1<\beta\leq \alpha \leq 2 \,.
\label{def:alphabetatheta}
\ee
Eq. (\ref{random_nu}) is, in general, driven by three scaling indices: 
(i) $\alpha$ and $\beta$, which are related to the shape of
the distribution, and (ii) $\phi$, i.e., the anomalous superdiffusive scaling 
of the MSD, related to the 
scaling exponent $\eta$ of the correlation function $R(t)$: $\phi = 2-\eta$,
$0 < \eta < 1$. 
The fundamental solution of the Space-Time Fractional Diffusion equation
(Section \ref{sec_stfde}, Supplementary Material), that is of particular interest for applications,
is obtained with the choice of parameters: $\phi = 2\beta/\alpha\ ;\quad 1 < 
\phi < 2$.
Interestingly, when $\phi \neq 2\beta/\alpha$,
Eq. (\ref{random_nu})
describes a {\it generalized space-time fractional diffusion} that is not
compatible with the Space-Time Fractional Diffusion equation.

\section{Numerical simulations}
\label{simulations}

\subsection{Simulation setup}

\noindent
In this section we carry out numerical simulations of the superdiffusive model
given by Eqs. (\ref{diff_eq1}-\ref{lang_free}) with random $\tau$ and $\nu$,
both to compare with analytical results and to verify the accuracy of our 
assumptions.
%
A total of $10\,000$ stochastic trajectories are computed for each simulation. 
To this goal, a statistical sample of $10\,000$ couples ($\tau$,$\nu$) is
firstly extracted by the respective distributions, each couple being associated
to one trajectory in the simulated ensemble.
In all simulations the following values are chosen: $\overline{\nu}=1$;
initial conditions $X_0=0$ and $V_0=0$ for all trajectories;
total simulation $T_{\rm sim}=10^3 \langle \tau \rangle$.

\vspace{.3cm}
Regarding the sampled populations of $\nu$ we consider 
three different distributions $f(\nu)$, corresponding to different
kinds of anomalous diffusion:
\begin{itemize}
\item[(1)]
{\it Gaussian anomalous diffusion with long-range correlations}:\\
A fixed value of $\nu$ is chosen to be equal for all trajectories.
This is a reduced model, whose 1-time PDF is given by Eqs.
(\ref{gauss_super_1},\ref{gauss_super_2},\ref{gauss_super_3}) and, for long
time lags, the stationary correlation function is given by 
Eq. (\ref{corr_free}) with $0<\eta<1$.
The only random parameter labeling the trajectories is the correlation 
time $\tau$. 
It is interesting to note that this model belongs to the class of 
Gaussian stochastic processes with stationary increments and long-range 
correlations, thus sharing the same basic features of FBM, but within a
completely different physical framework.
\item[(2)]
{\it Erd\'elyi--Kober fractional diffusion and Mainardi distribution}
\cite{pagnini-fcaa-2012,pagnini_etal-ijsa-2012}:\\ 
(parameter range: $\alpha=2$, $0 < \beta < 1$, $1 < \phi < 2$)
\be
\frac{1}{ \left( C \overline{\nu} t^\phi \right)^{1/2} } 
\frac12 M_{\beta/2} 
\left( \frac{x}{\left( C \overline{\nu} t^\phi \right)^{1/2} } \right)
 = \int_0^\infty {\cal G}(x,2 C \nu t^{\phi}) \frac{1}{\overline{\nu}} 
M_\beta
\left( \frac{\nu}{\overline{\nu}} \right) d\nu \ ,
\label{ek_fract}
\ee
being $M_ {\beta/2}/2 = K_{2,\beta}^0$; $M_\beta = K_{1,\beta}^{-1}$.
This is the solution of a fractional diffusion equation with Erd\'elyi--Kober
fractional derivative in time \cite{pagnini-fcaa-2012,pagnini_etal-ijsa-2012}.
%

For $\phi=\beta$ the solution of the Time Fractional Diffusion equation is 
recovered, i.e., Eq. (\ref{STFDE1}) of Supplementary Material with $\alpha=2$.
In this case the mean velocity diffusivity $\langle \nu \rangle$ is finite and
can be computed by applying the formula for the moments of $M_\beta$
\cite{mainardi_etal-fcaa-2001}:
\be
\langle \lambda^\delta \rangle = \int_0^\infty \lambda^\delta M_\beta(\lambda)
d\lambda = \frac{\Gamma(\delta+1)}{\Gamma(\beta\delta+1)}, \quad \delta>-1\ .
\label{moment_mainardi}
\ee
Thus:
\be
\langle \nu \rangle = \int_0^\infty \nu f(\nu) d\nu =
\int_0^\infty \frac{\nu}{\overline{\nu}} M_\beta \left( 
\frac{\nu}{\overline{\nu}} \right) d\nu = \frac{\Gamma(2)}{\Gamma(1+\beta)} 
\ {\overline{\nu}}
\label{mean_nu}
\ee
\item[(3)]
{\it Generalized Space Fractional Diffusion and extremal L\'evy distributions}:\\
(parameter range: $\beta=1$; $1 < \alpha <2$; $1 < \phi <2 $)
\be
\frac{1}{ \left( C \overline{\nu} t^\phi \right)^{1/2}  } 
L_\alpha^0 \left( \frac{x}{\left( C \overline{\nu} t^\phi \right)^{1/2} 
 } \right)
 = \int_0^\infty {\cal G}(x,2 C \nu t^\phi) \frac{1}{\overline{\nu}} 
L_{\alpha/2}^{-\alpha/2} \left( \frac{\nu}{\overline{\nu}} \right) d\nu \ ,
\label{space_fract}
\ee
where $L_\alpha^\theta$ is the L\'evy stable density of scaling $\alpha$ and
asymmetry $\theta$ and $L_\alpha^0 = K_{\alpha,1}^0$; 
$L_{\alpha/2}^{-\alpha/2} = K_{\alpha/2,1}^{-\alpha/2}$.
The moments of both PDFs $L_\alpha^0$ and $L_{\alpha/2}^{-\alpha/2}$
are not finite. In particular: $\langle\nu\rangle = \infty$.
For $\phi = 2/\alpha$ the solution of the Space Fractional 
Diffusion equation is recovered, i.e., Eq. (\ref{STFDE1}) of Supplementary
Material with $\beta = 1$.
\end{itemize}
%

For the random generation of $\nu$ we refer to the algorithms 
discussed and used in Ref. \cite{pagnini_etal-fcaa-2016} 
(Eq. (4.9) for the L\'evy extremal distribution and Eq. (4.6) for the Mainardi 
distribution), based on the Chambers--Mallows--Stuck algorithm for
the generation of L\'evy random variables
\cite{chambers_etal-jasa-1976,weron-spl-1996}.
The sampled population of $\tau$ is extracted from the PDF $g(\tau)$, 
Eq. (\ref{pdf_gtau}), using the numerical random generator described in 
Section \ref{appendix_num} of Supplementary Material. It is worth noting that
this algorithm is semi-analytical, 
that is, asymptotic solutions are used for both short and long $\tau$, while 
in the intermediate regime the algorithm is completely numerical.
The numerical scheme for the Langevin equation is described in the 
Supplementary Material, Section \ref{num_scheme}.

\subsection{Discussion of numerical results}
\label{results}

\noindent
Numerical simulations have been carried out for different values of scaling 
parameters and show qualitatively good agreement with analytical results for 
both ensemble averaged MSD $\sigma^2_X(t)$ and PDF $P(x,t)$. 
The goodness of comparison decreases as the parameters get closer to the 
extremal allowed values of the scaling parameters that are more far from 
standard and/or Markovian diffusion (i.e., $\eta=1$, $\alpha=2$, $\beta=1$).

\noindent
It is important to notice that, while the random generator of $\nu$ does 
not essentially determine any criticality in the numerical algorithm,
the role of the parameter $\tau$ in the numerical implementation of the model
is much more delicate.
This aspect is strictly related to the equilibrium properties of single
trajectories and of the overall system.
In fact, the derivation of our model
is based on the assumption of an equilibrium/stationary condition for all the
sample paths in the statistical ensemble.
This condition is exactly true only for $t=\infty$,
while, for whatever finite time $t$, is clearly well approximated only for
those trajectories satisfying the condition $\tau < t$. 
Conversely, due to the slow decaying power-law tail in the $g(\tau)$
distribution, relaxation times $\tau$ much longer than $\langle \tau \rangle$
have non-negligible probabilistic weights.
Thus, $\langle \tau \rangle$ does not really characterize the 
relaxation/correlation time of all stochastic trajectories, each one 
experiencing its own time scale to reach the equilibrium/stationary condition. 

\noindent
Then, two crucial aspects need to be verified:
does an equilibrium condition exists ?
Is the time scale to reach such equilibrium finite ?
%
%
%

\noindent
The working hypothesis to be checked is that, despite the inverse power-law
tail in $g(\tau)$,
the statistical weights of sufficiently large $\tau$ are negligible enough
to get a global equilibrium condition in the range $t\gg\langle\tau\rangle$.
This is a crucial aspect regarding the self-consistency of the model with
respect to the existence of a global stationary condition and, least but not
last, the comparison with experimental data.

\begin{figure}[!h]
\centering
\subfloat[][]
	{\includegraphics[width=.48\textwidth, height=0.26\textheight]{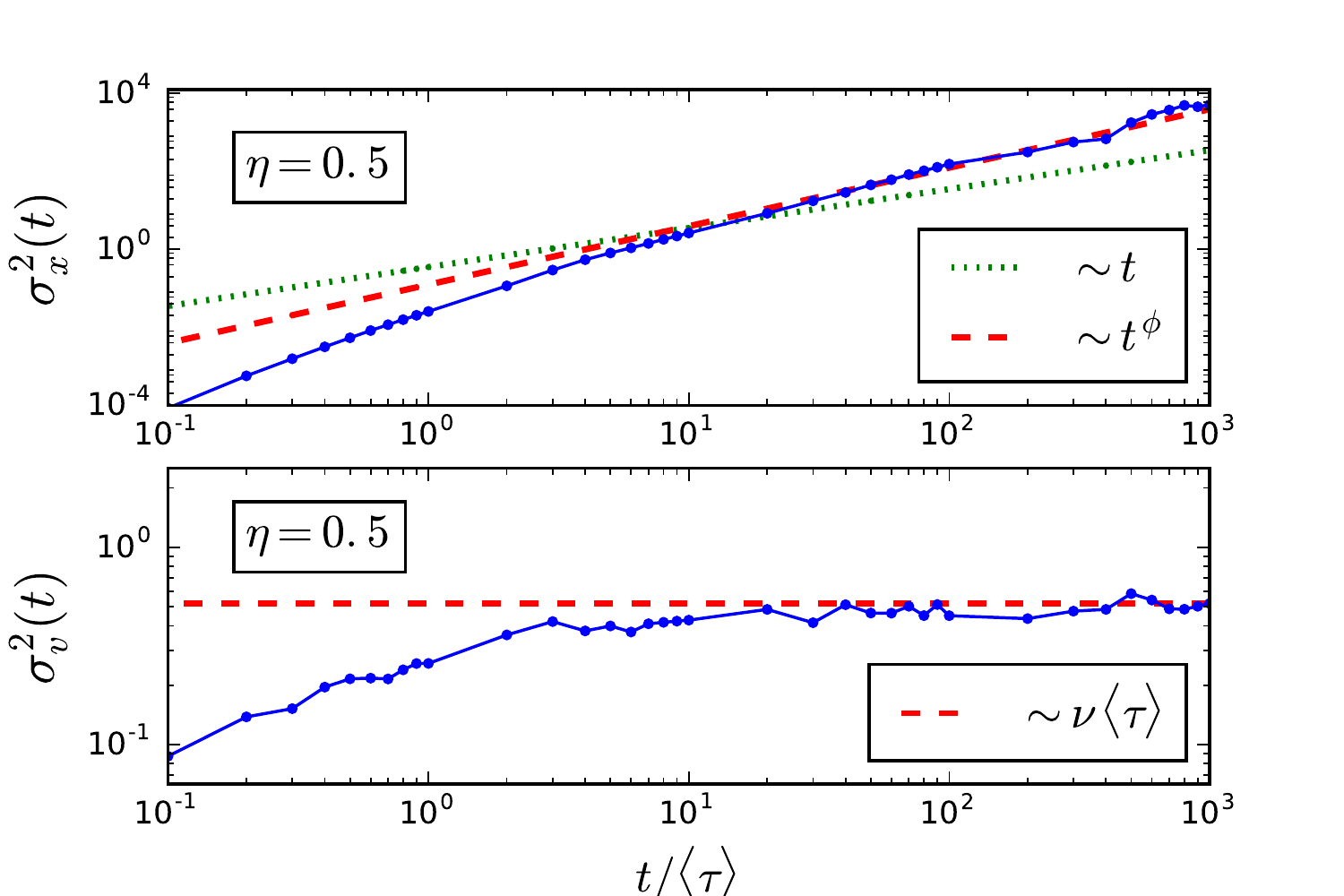}} \quad
\subfloat[][]
	{\includegraphics[width=.48\textwidth, height=0.26\textheight]{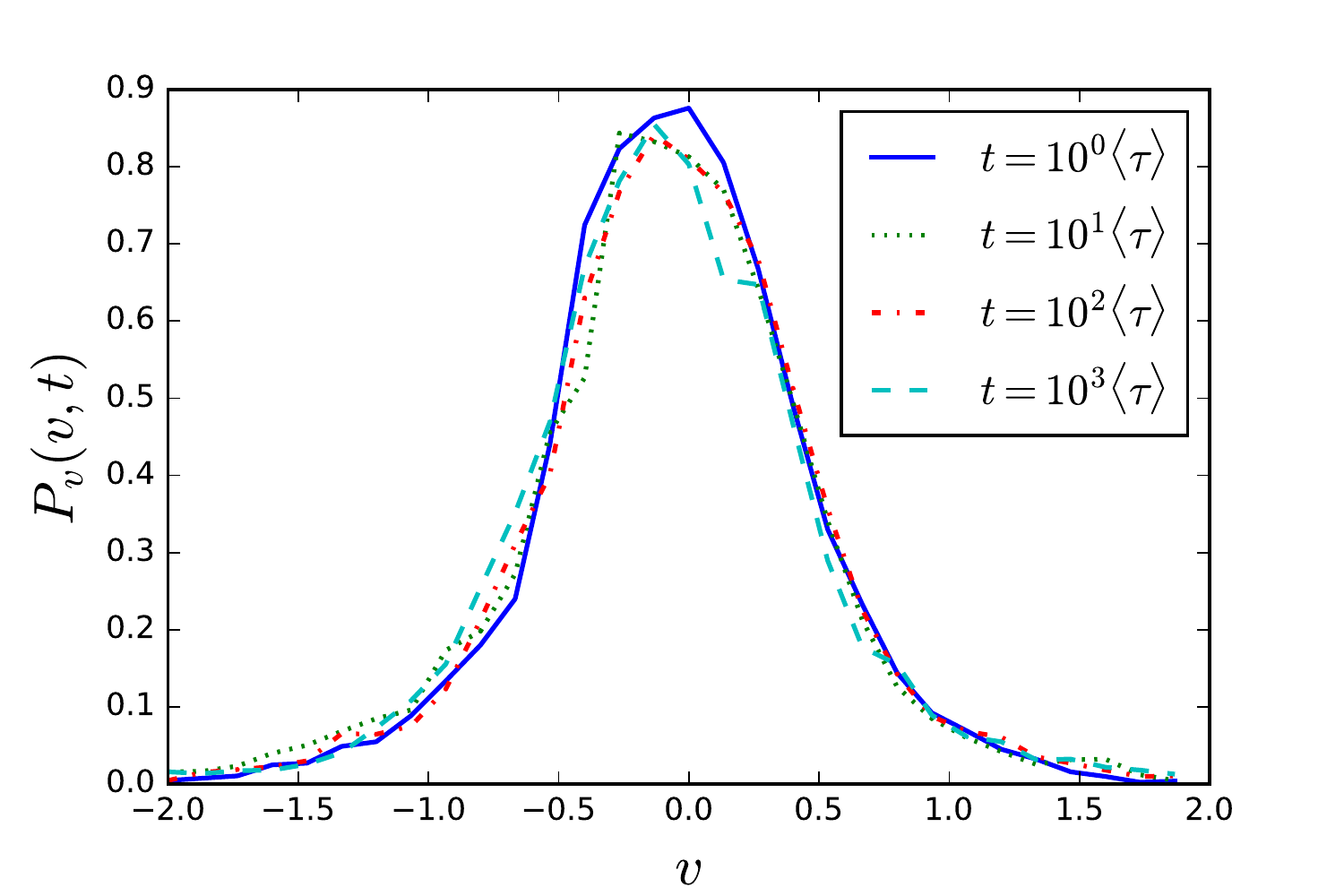}}\\
\caption{ 
  Superdiffusion with $\eta=0.5$ ($\phi=2-\eta=1.5$, $\la\tau\ra=0.52$) and
  fixed $\nu=1$ (Gaussian case).
  (a) MSD of velocity (bottom panel) and position (top panel); (b) velocity 
  (Gaussian) PDF $P(x,t)$ at different times.
 }
\label{super_img}
\end{figure}

\noindent
The numerical simulations proved that a (global) stationary state indeed exists
and that the equilibrium condition and the expected anomalous
diffusion regime in the MSD are reached for times sufficiently larger than
$\la\tau\ra$.
%
%
In Fig. \ref{super_img} we show the results for the simulation of a statistical
sample of $10\,000$ trajectories with $\eta=0.5$ and fixed $\nu=1$ (Gaussian 
case).
From bottom panel (a) and panel (b) it is clear that the system reaches the 
stationary state within a time of the order $t \approx 10 \la\tau\ra$ or less,
which is the time the particle needs to reach the theoretical stationary
velocity variance $\la V^2 \ra_{\rm st} = \la\nu\ra\la\tau\ra$ (bottom panel (a)) 
an the long-time diffusive scaling $\phi=2-\eta=1.5$ (top panel (a)).
From panel (b) it is clear that velocity fluctuations reached a 
stationary/equilibrium condition.
This characteristic time depends on $\eta$ as it decreases while $\eta$ 
increases. This feature is due to $g(\tau)$
that, for $\eta$ approaching 1, becomes more and more peaked tending
towards a Dirac $\delta$ function. For $\eta=1$ a unique value of $\tau$ is 
chosen for all particles, so that the relaxation time of the whole system 
becomes $\tau$ itself and we fall back into standard diffusion. 
Thus, numerical simulations show that the stationary condition is reached at 
reasonable (i.e., not too much large) times. 
This is a good indication that the model can well compare with experimental
data, anomalous diffusion emerging in a given temporal range that is not
too short neither too long.
This is true for values of scaling indices that are not too close
to extremes of the definition interval (e.g., $\beta$ far from $0$), except 
those extremal values corresponding to time and space locality, i.e., standard 
diffusion and/or Markovian processes.

\begin{figure}[!h]
\centering
\subfloat[][]
	{\includegraphics[width=.48\textwidth, height=0.26\textheight]{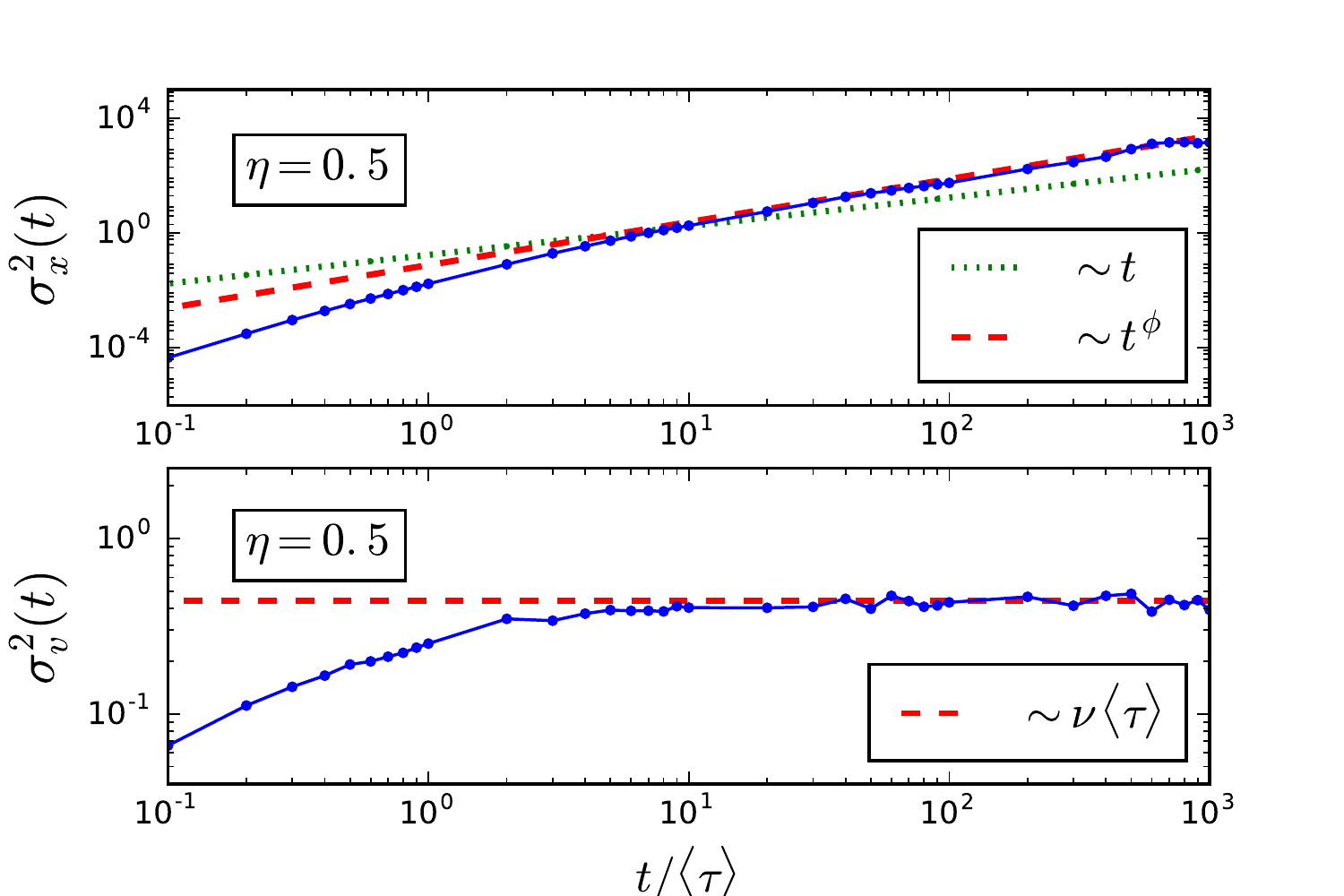}} \quad
\subfloat[][]
	{\includegraphics[width=.48\textwidth, height=0.26\textheight]{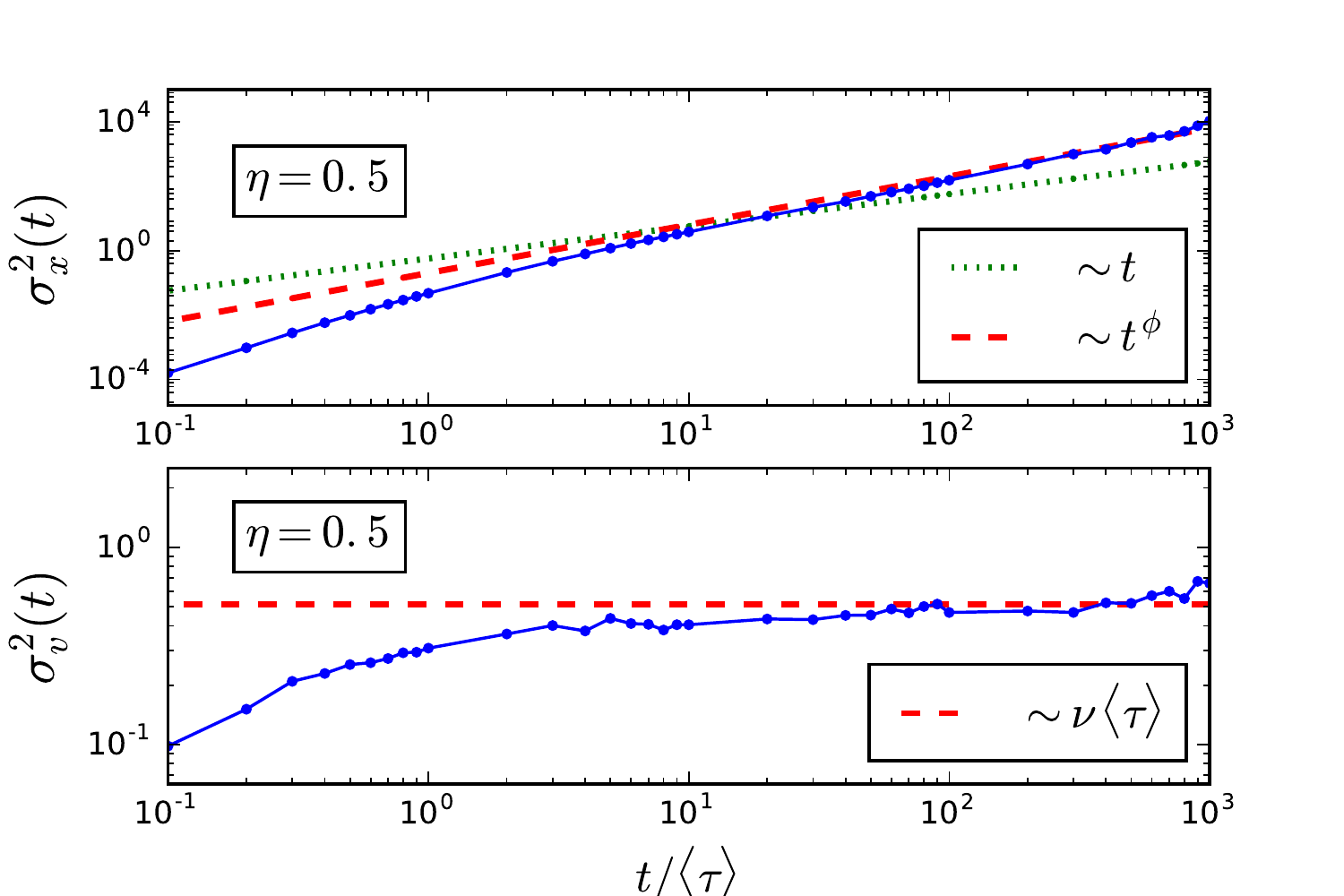}}\\
\caption{ 
  Superdiffusion with $\eta=0.5$ ($\phi=2-\eta=1.5$, $\la\tau\ra=0.52$) and 
  fixed $\nu=1$ 
  (Gaussian case). MSD of velocity (bottom panels) and position (top panels)
  (a) Sampled set with $\la\tau\ra = 0.44$ and $\tau_{\rm max} = 75.2$; 
  (b) Sampled set with $\la\tau\ra = 0.66$ and $\tau_{\rm max} = 1580.7$.
 }
\label{tausets_img}
\end{figure}

\noindent
In the case of inverse power-law tails, different statistical samples extracted
from the distribution $g(\tau)$ can have quite different statistics 
(e.g., different $\la\tau\ra$).
Due to the slow power-law decay and the unavoidable finiteness of the 
statistical sample, the maximum value $\tau_{\rm max}$ can also vary
significantly among different samples.
Numerical simulations for five different sampled sets of $\tau$ are carried 
out with $\la\tau\ra = 0.52, 0.44, 0.5, 0.46, 0.66$ and 
$\tau_{\rm max} = 279.2, 75.2, 91.9, 200.4, 1580.7$.
The simulations are found to be well comparable with each other. 
This can be seen in Fig. \ref{tausets_img}, where we compare the two sampled 
sets of $\tau$ having the minimum and maximum values of $\tau_{\rm max}$ 
(Gaussian model).
Even if these values are different by orders of magnitude (from $75.2$ to 
$1580.7$), the dependence on $\tau_{\rm max}$ is weak, as the time to reach 
stationarity changes from about $10-30$ to $60-80$ (see the velocity variances
in the bottom panels).
Further, the time to reach the stationary state does not change when comparing 
the Gaussian model with non-Gaussian ones (random $\nu$).


%
%

\vspace{.3cm}
\begin{figure}[!h]
\centering
	{\includegraphics[width=.6\textwidth, height=0.26\textheight]{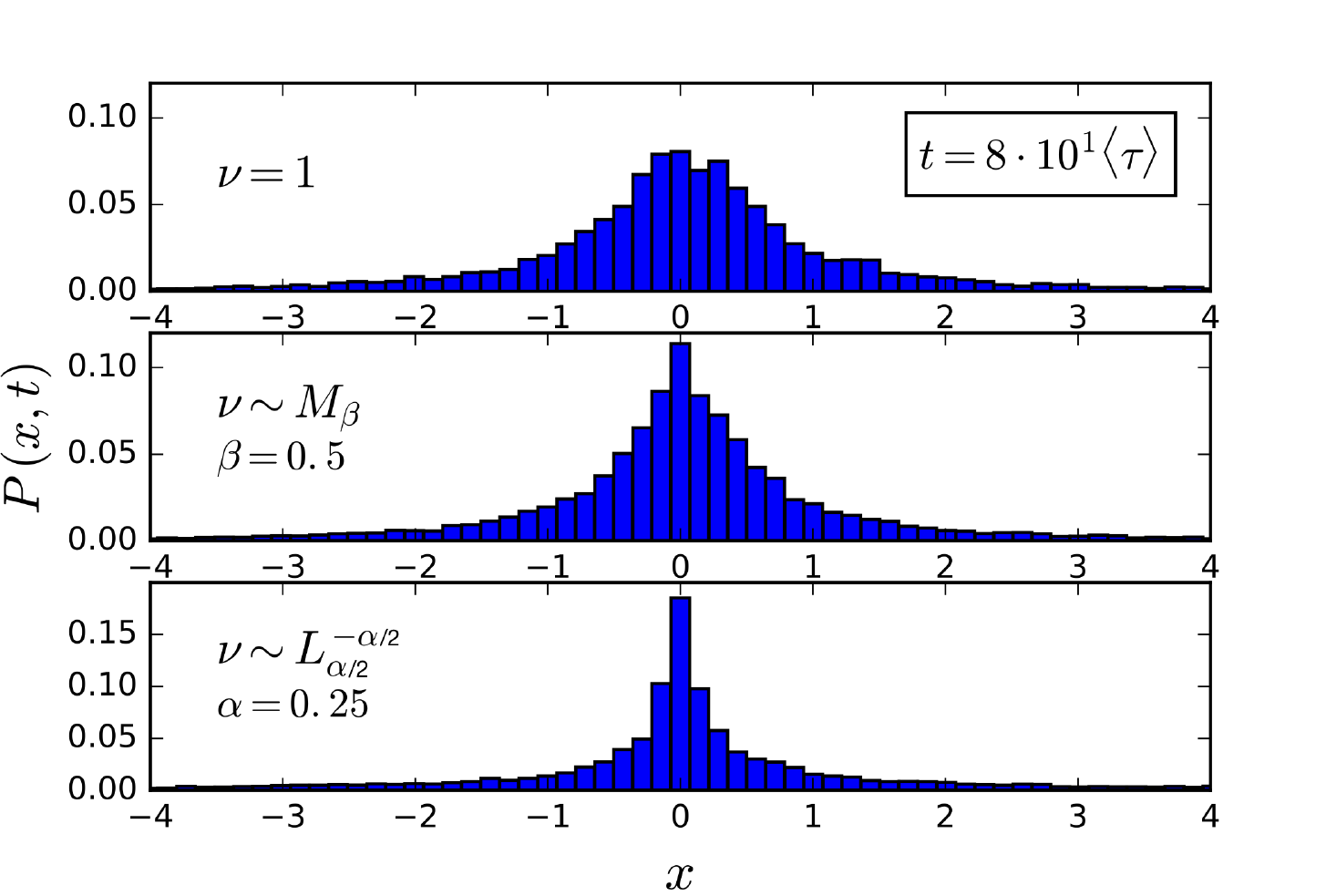}} \quad
\caption{ 
  Superdiffusion with $\eta=0.5$ ($\phi=2-\eta=1.5$, $\la\tau\ra=0.52$). 
  Comparison of PDFs $P(x,t)$ for different distributions $f(\nu)$. 
  Top panel: $\nu$ fixed, i.e., $f(\nu)=\delta(\nu-\overline{\nu})$ with 
  ${\overline \nu} = 1$ (Gaussian case); 
  intermediate panel: $M_\beta(\mu)$ distribution, $\beta=0.5$ (Erd\'elyi--Kober 
  fractional diffusion); bottom panel:  $L_{\alpha/2}^{-\alpha/2}$ distribution, 
  $\alpha=0.5$ (Generalized space fractional diffusion).
  }
\label{super_img_1}
\end{figure}
\begin{figure}[!h]
\centering
\subfloat[][]
	{\includegraphics[width=.48\textwidth, height=0.26\textheight]{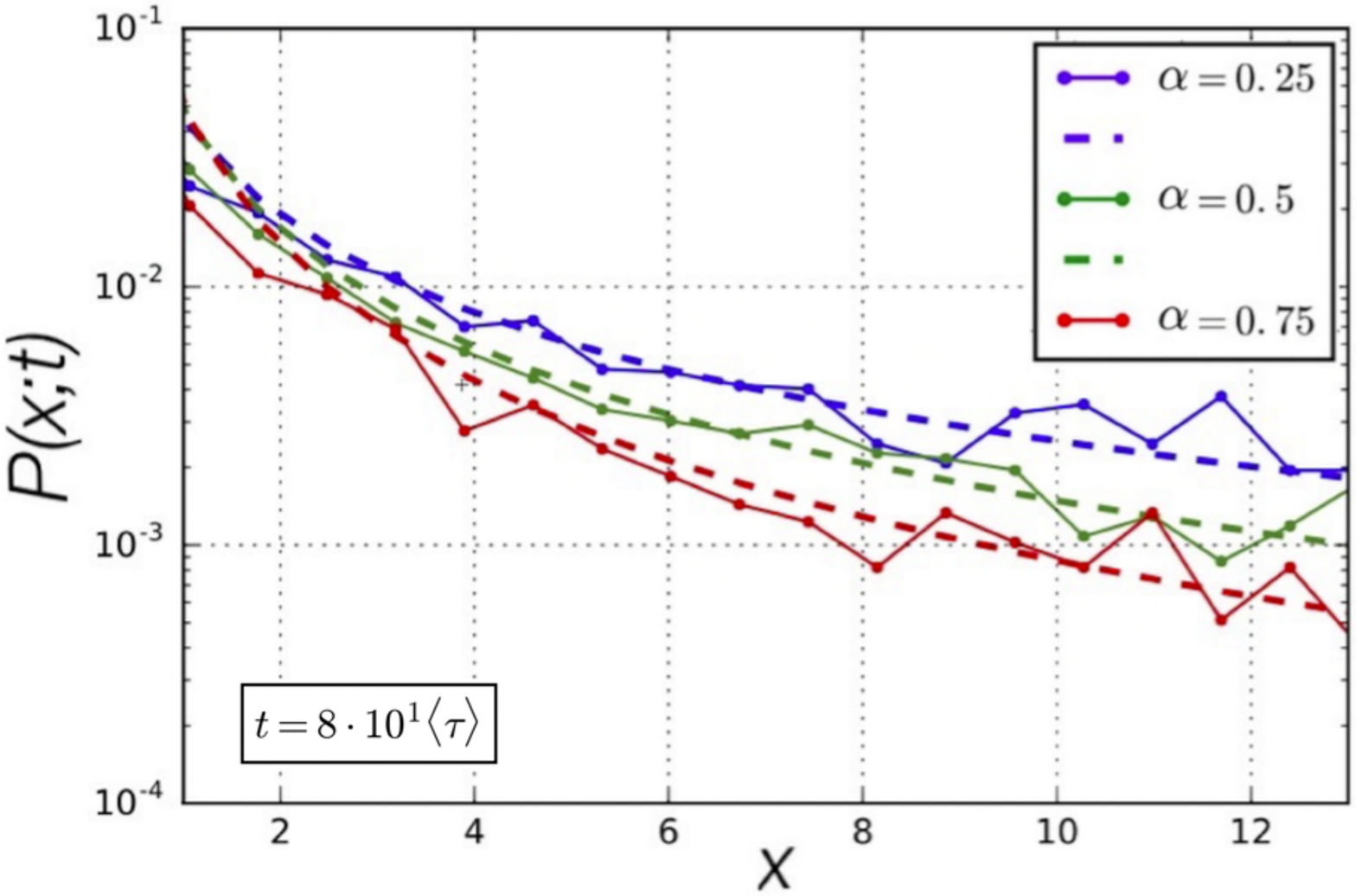}} \quad
\subfloat[][]
	{\includegraphics[width=.48\textwidth, height=0.26\textheight]{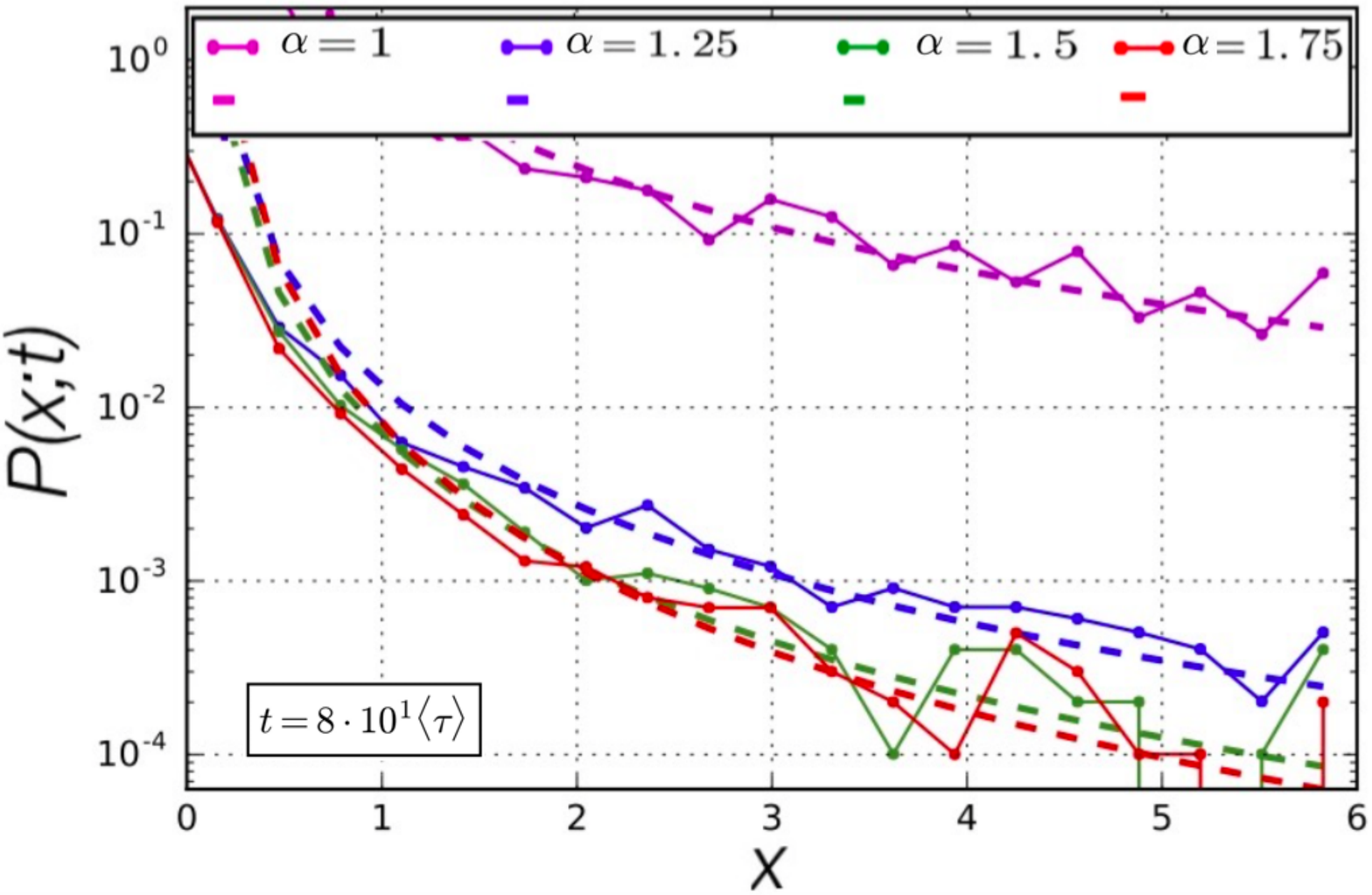}}\\
\subfloat[][]
	{\includegraphics[width=.31\textwidth, height=0.17\textheight]{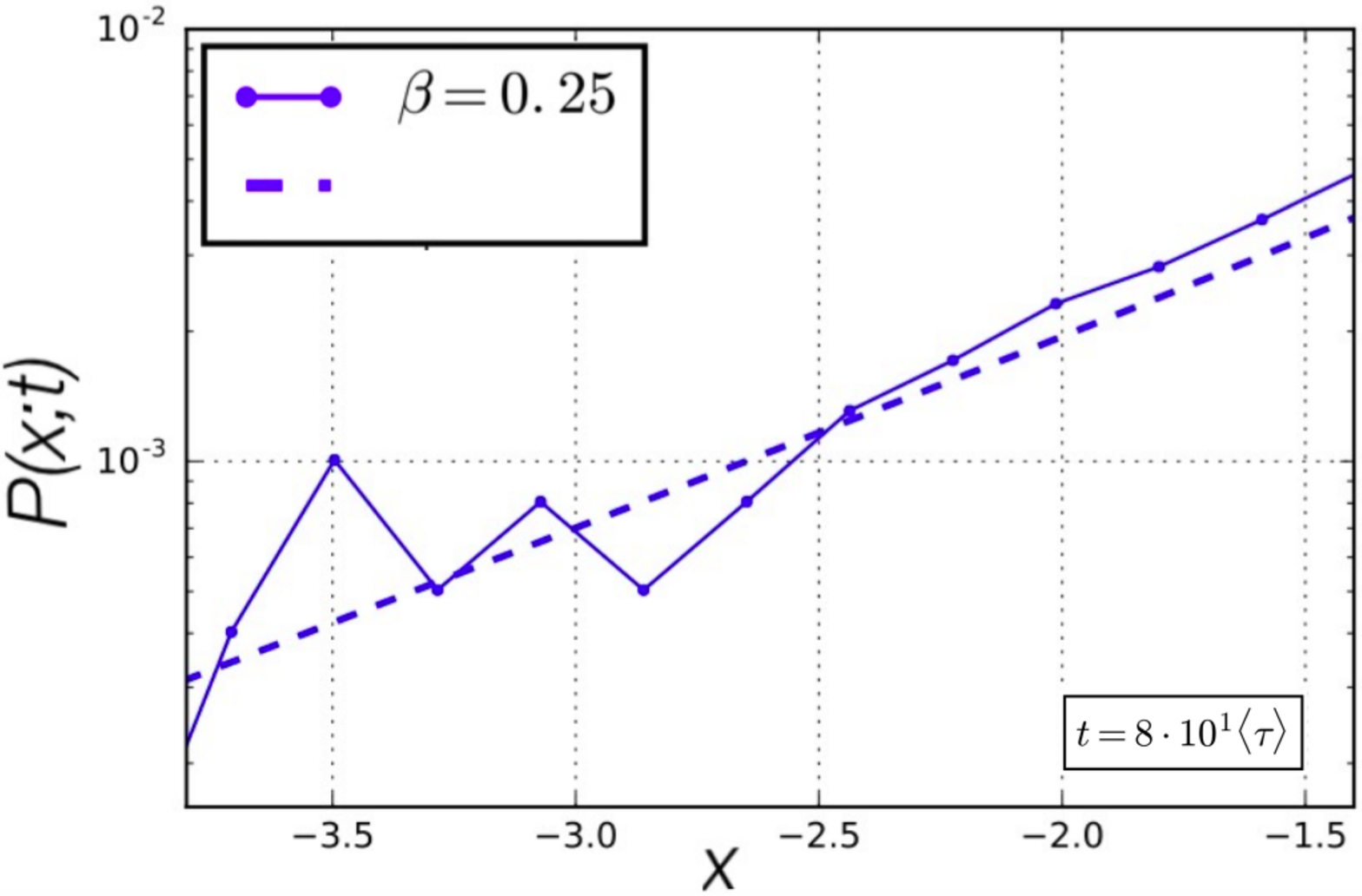}}\quad
\subfloat[][]
	{\includegraphics[width=.31\textwidth, height=0.17\textheight]{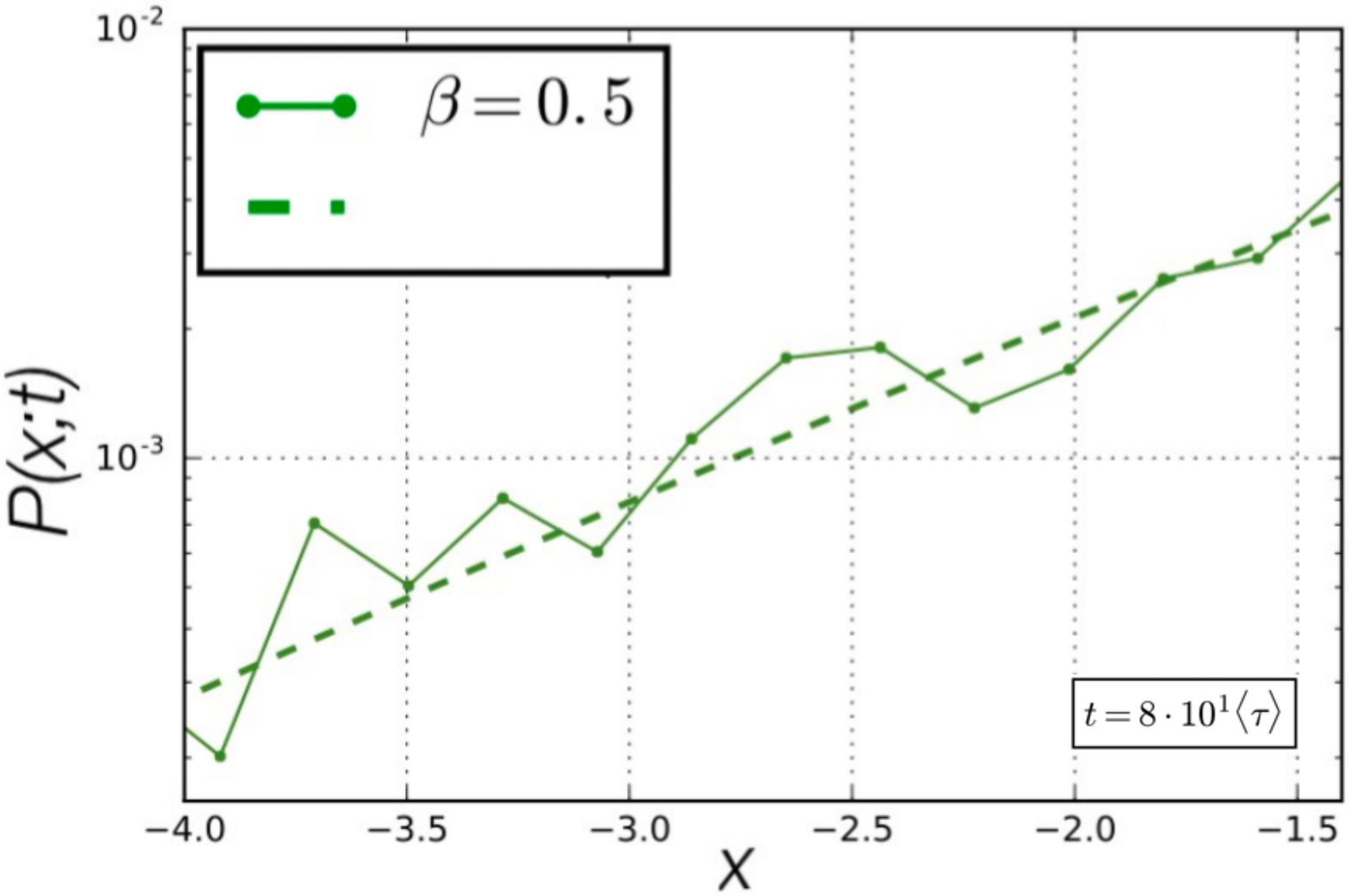}}\quad
\subfloat[][]
	{\includegraphics[width=.31\textwidth, height=0.17\textheight]{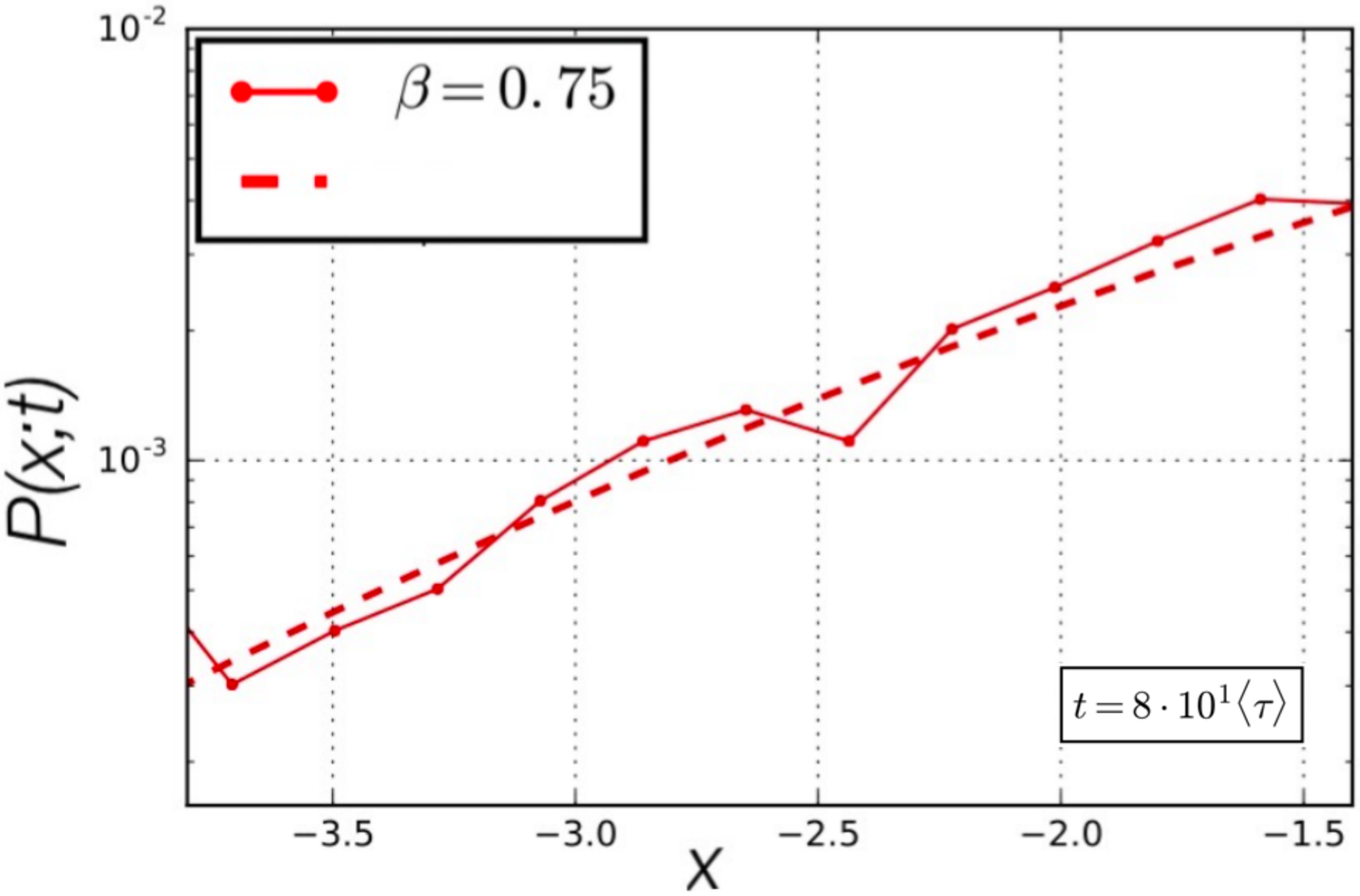}} \\
\caption{
  Superdiffusion with $\eta=0.5$ ($\phi=2-\eta=1.5$, $\la\tau\ra=0.52$). 
  Comparison of analytical
  and numerical position PDFs $P(x,t)$ in the asymptotic regime. Top panels: 
  different values of the scaling index $\alpha$ ($L_{\alpha/2}^{-\alpha/2}$, 
  Generalized space fractional equation); bottom panels: different values of 
  the scaling index $\beta$ ($M_{\beta}$, Erd\'elyi--Kober fractional diffusion).
 }
\label{super_img_2}
\end{figure}

\noindent
Fig. \ref{super_img_1} qualitatively shows the changes in the shape of the 
position PDF $P(x,t)$ due to the $\nu$ randomization. The top panel displays a typical Gaussian shape.
Finally, in Fig. \ref{super_img_2} we compare the asymptotic tails of 
analytical solutions for the position PDF $P(x,t)$ with the corresponding
histograms computed from numerical simulations.
The comparison, carried out for $\eta=0.5$, show a good agreement for
all the used values of $\alpha$ and $\beta$. Similar agreement was seen in 
simulations, not shown here, that were carried out for $\eta=0.25$ and
$\eta=0.75$.

\section{Concluding remarks}
\label{conclusion}

\noindent
We have introduced and discussed a novel modeling approach based
on a linear Langevin equation (friction-diffusion process) driven by a 
population of two parameters: relaxation time $\tau$ and
{\bf velocity }
diffusivity $\nu$, with distributions properly chosen to get anomalous diffusion
(Gaussian or fractional).
It is worth noting that both $\tau$ and $\nu$ directly characterize the
velocity's dynamics and only indirectly the position dynamics. In particular,
$\nu$ determines the diffusion properties of velocity and, for normal diffusion,
its dimensional units are $[\nu]=[V^2]/[T]$.
Gaussian anomalous diffusion is obtained by considering a constant
{\bf velocity } diffusivity
and imposing the correct power-law correlation function compatible with
MSD anomalous scaling.
Fractional diffusion is derived by imposing the particular PDFs that are
fundamental solutions of EKFDE or STFDE.
Our stochastic model can also generate a {\it generalized fractional diffusion},
whose more general expression for the 1-time PDF is given in Eq. 
(\ref{random_nu}). In this PDF the space-time scaling relationship is not
related to the scaling indices defining the shape of the PDF itself, as in 
the fractional diffusion.
%
%

%
%
At variance with other HDPs, the inclusion of viscosity in our model allows us to include the
effect of relaxation.
The distribution of relaxation times $\tau$ is then a crucial property that is
here derived by imposing the emergence of anomalous diffusion, retaining at the
same time the Gaussianity and stationarity of velocity increments.

Another interesting aspect is the weak ergodicity breaking established in biological motion
data \cite{burov_etal-pccp-2011,he_etal-prl-2008,schulz_etal-prl-2013} and defined by the
inequality of ensemble and time averaged MSD in anomalous diffusion processes. 
%
In particular, even if the ensemble averaged MSD is given by Eq. 
(\ref{anom_diff}), the time averaged MSD depends linearly on the time lag.
%
%
In the model here proposed, the single trajectory is driven by the linear 
Langevin equation describing the Ornstein--Uhlenbeck process, which is 
characterized by the crossover between a short-time ballistic diffusion: 
$\sigma^2_X(t) \sim t^2$; and a long-time standard (Gaussian) diffusion: 
$\sigma^2_X(t) \sim t$.
Thus, the single trajectory naturally follows a standard diffusion law in the
long-time limit.
The non-ergodic behavior is modelled by considering the randomness of
physical properties and, in particular,  relaxation time
and velocity diffusivity, the first one driving the drift 
(linear viscous drag) and the second one driving the noise, respectively.

An important observation regarding the comparison between our ggBM-like modeling
approach and other similar approaches is in order.
All these heterogeneity--based models attempt to describe the role of
heterogeneity in triggering the emergence of long-range correlations and
anomalous diffusion.
However, superstatistics and other models (fluctuating friction or mass, DDMs)
lie on mimicking heterogeneity through the temporal stochastic dynamics or
modulation of some parameters driving the particle's dynamics.
On the contrary, ggBM-like models explicitly describe the heterogeneity as
inter-particle fluctuations of parameters that are responsible for long-range
correlations, in agreement with approaches based on polydispersity where
classical thermodynamicsholds \cite{gheorghiu_etal-pnas-2004}.

%
%

Future investigations are needed not only to better understand these last
observations but also, on one side, to characterize our 
proposed model in terms of several statistical indicators that are commonly 
used in the analysis of biological motions and,
on the other side, to better understand the link of the parameter distributions
to the observable physical properties of the complex medium.
Finally, our modeling approach can be extended to the subdiffusive case by
considering a kind of trapping mechanism such as a stable fixed point.

\vspace{.5cm}
\noindent
{\bf Acknowledgements}

\small
This research is supported by the Basque Government through the BERC 2014--2017
and BERC 2018--2021 programs,  and by the Spanish Ministry of Economy and
Competitiveness MINECO through BCAM Severo Ochoa excellence accreditation
SEV--2013--0323 and through project MTM2016--76016--R ''MIP''. 
VS acknowledges BCAM, Bilbao, for the financial support to her internship
research
period during which she developed her Master Thesis research useful for her
Master degree in Physics at University of Bologna, 
and SV acknowledges the University of Bologna for the financial support through
the ''Marco Polo Programme'' for her PhD research period abroad spent at
BCAM, Bilbao, useful for her PhD degree in Physics at University of Bologna.
PP acknowledges financial  support from Bizkaia Talent and European Commission
through COFUND scheme, 2015 Financial Aid  Program for Researchers, project
number AYD--000--252 hosted at BCAM, Bilbao.

\normalsize

\section*{References}




%










\newpage

\section*{Supplementary Material}

%
%



%

\section{General condition for the emergence of anomalous diffusion}
\label{random_media}

\noindent
Diffusion is described through the following simple, but general stochastic
equation:
\be
\frac{dX_t}{dt} = V_t\
\label{diff_eq}
\ee
being $V_t$ a stochastic process describing a generic random fluctuating signal.
Here $X_t$ and $V_t$ are the position and velocity of a
particle moving in a random medium, respectively.
For a generic, nonstationary process, the two-time 
Probability Density Function (PDF) $p(V_1,t_1;V_2,t_2)$ depends on both times
$t_1$ and $t_2$.
Similarly, the correlation function 
$$
\la V_{t_1} \cdot  V_{t_2}\ra = \int V_1 \ V_2 \ p(V_1,t_1;V_2,t_2) dV_1 dV_2
$$ 
is, in general, a function of the times $t_1$ and $t_2$\footnote{
This also means that the statistics of $V_t$ increments: 
$\Delta V_{t_1,t} = V_{t_1+t} - V_{t_1}$, depend not only on the time lag 
$t$, but also on the initial time $t_1$
}.

Now, by integrating in time the above kinematic equation (\ref{diff_eq}), 
making the square and the ensemble average, we get the Mean Square
Displacement (MSD):
\be
\sigma_X^2(t) = \la (X_t-X_0)^2 \ra = \int_{0}^t dt^\prime \int_{0}^t 
dt^{\prime\prime} \la V_{t^\prime} \cdot V_{t^{\prime\prime}} \ra \ ,
\label{diff_corrv}
\ee
where, in order to get : $\langle X_t \rangle = X_0$, we assumed a uniform 
initial position $X_0$.
%
In the stationary case, the two-time statistics, including the correlation 
function, depends only on the time lag $t = |t_1-t_2|$, and the above formula 
reduces to:
\be
\sigma_X^2(t) = \int_{0}^t dt^\prime \int_{0}^t 
dt^{\prime\prime} R(|t^\prime-t^{\prime\prime}|) = 
2 \int_{0}^t (t - s) R(s) \, ds \ ,
\label{taylor21_1}
\ee
or, equivalently:
\be
\frac{d\sigma_X^2(t)}{dt} = 2 \int_{0}^t  R(s) ds \ .
\label{taylor21_2}
\ee
where $R(t) = \la V_{t_1+t} \cdot V_{t_1}\ra = \la V_t \cdot V_0\ra$ is the
stationary correlation function.
Notice that these expressions have very general validity, independently of the
particular statistical features of $V_t$.

These expressions were firstly published by Taylor in 1921 \cite{taylor1921},
which implicitly formulated the following:

\vspace{.3cm}
{\bf Theorem (Taylor 1921)}

Given the stationary correlation function $R(t)$, let us define the 
correlation time scale:
\be
\tau_c=\int_0^\infty \frac{R(s)}{R(0)} ds\ ,\quad 
R(0) = \langle V^2 \rangle_{\rm st} \ .
\label{tau_corr}\footnote{
Notice that the variance $\langle V^2 \rangle_{\rm st}$, being a one-time
statistical feature, is a constant in the stationary case.
}
\ee
Then, if the following condition occurs:
\be
0 \ne \tau < +\infty \ ,
\label{finite_tauc}
\ee
{\it normal diffusion} always emerges in the long-time regime:
\be
t\gg \tau_c \quad \Rightarrow \quad \sigma_X^2(t) = 2 D_{_X} t\ ,
\label{norm_diff_1}
\ee
thus defining the long-time spatial diffusivity $D_{_X}$:
\be
D_{_X} := \frac12 \lim_{t \rightarrow +\infty}\frac{d\sigma_{\rm x}^2}{dt}(t)
\label{norm_diff_2}
\ee
independently from the details of the microdynamics driving the
fluctuating velocity $V_t$.

It is worth noting that, substituting Eq. (\ref{taylor21_2}) into
Eq. (\ref{norm_diff_2}) and using $R(0) = \langle V^2 \rangle_{\rm st}$ 
(Eq. (\ref{tau_corr})), we get:
\be
D_{_X} = \tau_c \ \langle V^2 \rangle_{\rm st}\ ,
\label{einstein_smo_taylor}
\ee
which is a general form of the Einstein--Smoluchovsky relation
\cite{einstein_adp05}\footnote{
Interestingly, this relation is here derived in a very general framework, i.e.,
for a generic fluctuating signal $V_t$, with the only assumption of the
existence of a stationary regime in the long-time limit. 
As known, the stationary condition usually emerges in correspondence of 
motion reaching an equilibrium state. However, the stationary condition
is more general with respect to equilibrium and, for this reason, we prefer to 
leave the notation ``${\rm st}$'' for "stationary'' instead of ``${\rm eq}$'' 
for "equilibrium''.
}.

\vspace{.3cm}
Taylor's theorem gives in Eq. (\ref{finite_tauc}) the general conditions to 
get normal diffusion, i.e., a
linear scaling in the variance: $\langle X^2\rangle\sim t$).
This result has a very general validity, independently from the statistical 
features of the stochastic process $V_t$.
The theorem also establishes the regime of validity of normal diffusion, 
given by the asymptotic condition $t\gg\tau_c$.
As a consequence, the emergence of {\it anomalous diffusion} is strictly 
connected to the failure of the assumption (\ref{finite_tauc}). 
In particular, we get two different cases:
\begin{itemize}
\item
{\it Superdiffusion}:
\be
\tau_c = \infty:\ \langle X^2\rangle\sim t^\phi\ \ {\rm with}\ \ 1 < \phi \le 2 
\ \ {\rm or}\ \ \langle X^2\rangle = \infty\ .
\label{superdiff}
\ee
\item
{\it Subdiffusion}:
\be
\tau_c = 0:\ \langle X^2\rangle\sim t^\phi\ \  {\rm with}\ \  0 < \phi \le 1 \ .
\label{subdiff}
\ee
\end{itemize}
In order to get $\tau_c=0$ and, thus, subdiffusion, velocity
anti-correlations must emerge. This means that there exist time lags $t$ such 
that $R(t)<0$ 
(e.g., the anti-persistent Fractional Brownian Motion, with $H<0.5$). 
Being $R(0)=\langle V^2 \rangle_{\rm st} > 0$, in subdiffusion the correlation 
function is surely positive in the short-time regime and (i) becomes negative 
in the long-time regime or (ii) oscillates between positive and negative 
values\footnote{
A correlation time scale, different from the above definition of
$\tau_c$ can be sometimes introduced for subdiffusion (e.g., the time period 
in a harmonic correlation function), but it does not have the 
meaning of discriminating a long-time regime with normal diffusion from 
a short-time regime.
}.

The failure of Taylor's theorem and of condition (\ref{finite_tauc}) is the 
main guiding principle exploited here to derive stochastic models 
for anomalous diffusion.


\subsection{Application to Fractional Brownian Motion}

\begin{figure}
\includegraphics[width=0.50\textwidth, height=0.80\textwidth,angle=270]{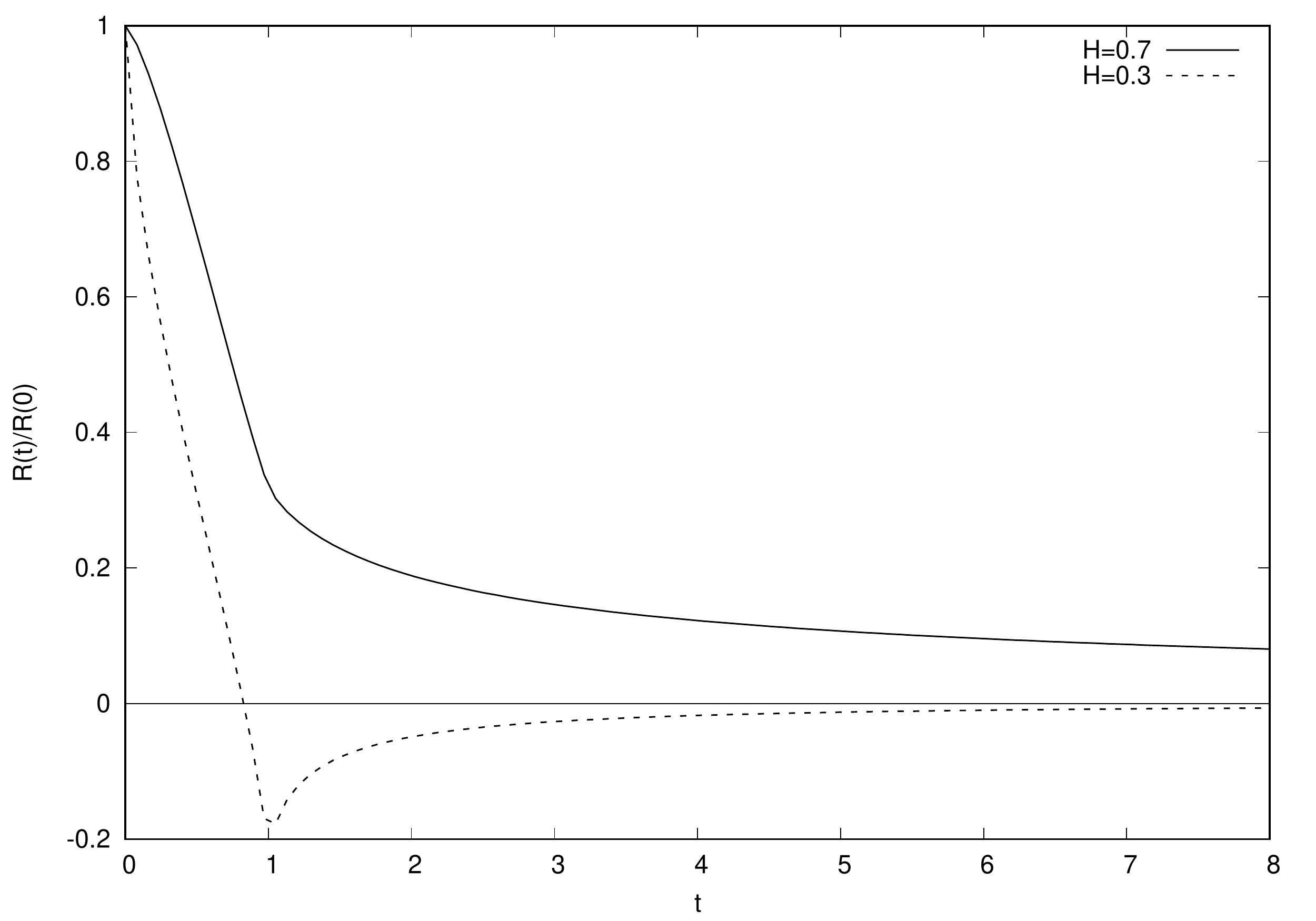}
\caption{\label{fig_fbm_corr} Autocorrelation function of FBM increments
  $V_1(t)$: $R(t)=\la V_1(s)V_1(s+t) \ra$: persistent ($H=0.7$) vs.
  antipersistent ($H=0.3$) case.
}
\end{figure}

\noindent
The Fractional Brownian Motion (FBM) $B_H(t)$ was introduced by Mandelbrot and
Van
Ness in their famous 1968's paper \cite{mandelbrot_etal-siamrev-1968}.
Since then, thousands of papers have been devoted to both theoretical
investigations and applications of FBM (see, e.g., \cite{dieker_phd_2004}
for a review).
FBM is a Gaussian process with self-similar stationary increments and
long-range correlations. In formulas, FBM has the following properties:
\begin{itemize}
\item
    $B_H(t)$ has stationary increments;
\item
  $B_H(0)=0$; $\la B_H(t)\ra = 0$ for $t \ge 0$;
\item
  $\la B^2_H(t)\ra = t^{2 H}$ for $t \ge 0$;
  \item
    $B_H(t)$ has a Gaussian distribution for $t > 0$;
  \item
    the correlation function is given by:
    \be
    \la B_H(t) B_H(s) \ra = \frac{1}{2}\left\{t^{2 H}+s^{2 H}-|t-s|^{2 h} \right\}
    \label{fbm_corr}
    \ee
\end{itemize}
The FBM increments are given by:
\be
V_{\delta t} (s) = B_H(s+\delta t) - B_H(s) \ .
\nonumber
\ee
The process $V_{\delta t}(s)$ is also called
{\it fractional Gaussian noise}\footnote{
  This can be considered as a kind of velocity for the FBM, even if it must
be kept in mind that FBM, such as standard Brownian motion, does not have a
smooth velocity. In any case, the above considerations about velocity and
position and their statistical relationship can here be applied by substituting
velocity with the fractional Gaussian noise, i.e., the FBM increments over
a finite time step $\delta t$.
}.
Both $B_H(t)$ and $V_{\delta t}(s)$ are self-similar stocastic processes but,
at variance with $B_H(t)$, the increments $V_{\delta t}(s)$ are also stationary,
i.e., their statistical features do not depend on $s$, but only on $\delta t$.
$V_{\delta t}(s)$ is a Gaussian process and is uniquely defined by the mean,
variance and correlation function, which are derived from the above
listed properties of FBM:
\be
\la V_{\delta t}(s) \ra = 0 \ ;\quad \la V^2_{\delta t}(s) \ra = (\delta t)^{2 H}
\ee
\be
R(t) = \la V_{\delta t}(s) V_{\delta t}(s+t) \ra = 
\frac{1}{2}\left\{|t+\delta t|^{2 h} - 2 t^{2 H} + |t-\delta t|^{2 H}  \right\}
\label{corr_fbm_incr}
\ee
Then, we can say that FBM is a Gaussian process with stationary and self-similar
increments $V_s (\delta t)$, while FBM is Gaussian, self-similar but
not stationary. Eq. ~\eqref{corr_fbm_incr} also shows that, with the
exception of the standard Brownian motion ($H=1/2$), increments
$V_{\delta t}(s)$ are not independent each other.
Fractional Gaussian noise and FBM are exactly self-similar, i.e., they satisfy
the relationship:
$X(a t) = a^H X (t)$, the increment $V_{1}(s)$ with $\delta t = 1$ is usually
considered in both theoretical and experimental studies, as a generic $\delta t$
can be obtained by simply rescaling the process with the self-similarity
relationship.
In Fig. \ref{fig_fbm_corr} the increment correlation functions of a persistent
($H>0.5$) and of an antipersistent ($H<0.5$) FBM are compared.
It is evident that antipersistent FBM is associated with anticorrelations,
and this is the reason why subdiffusion emerges in this case.

\noindent
The asymptotics of the correlation function are easily obtained
by rewriting it in the following way (see \cite{dieker_phd_2004}, pages 6-7):
\be
R(t) =  \frac{1}{2} t^{2 h} h_H\left( \frac{\delta t}{t} \right)\ ,
\ee
being, for $x=\delta t/t < 1$:
\be
h_H\left( x \right) = (1+x)^{2 H}-2+(1-x)^{2 H}\ .
\label{aux_function}
\ee
The limit $t \rightarrow \infty$ corresponds to $x \rightarrow 0$ and the
Taylor expansion of $h_H\left( x \right)$ gives:
\be
h_H(x) = 2H (2H -1) x^2 + O(x^4)\ ,
\label{aux_func_taylor}
\ee
so that \cite{mandelbrot_etal-siamrev-1968,dieker_phd_2004}:
\be
R(t) \simeq H(2H-1)(\delta t)^2 t^{2H-2}\ .
\ee
Regarding the correlation time $\tau_c$ defined in Eq. ~\eqref{tau_corr}, we
can exploit the same asymptotic expansion used for $R(t)$.
Firstly, we apply Eq. ~\eqref{tau_corr} to a finite time $t$:
\be
\tau_c(t) =\int_0^t \frac{R(s)}{R(0)} ds\ ,\quad 
R(0) = \langle V^2 \rangle_{\rm st} \ ,
\ee
so that: $\tau_c = \lim_{t\rightarrow\infty}\tau_c(t)$.
Then, for the fractional Gaussian noise we get:
\be
\tau_c(t) = \frac{\delta t}{4H+2}\left\{\left(1 + \frac{t}{\delta t}
\right)^{2H+1} - 2 \left( \frac{t}{\delta t} \right)^{2H+1} +
\left| \frac{t}{\delta t} -1\right |^{2H+1}
\right\}\ .
\label{tau_c_finite}
\ee
Analogously to $R(t)$, this can be written as:
\be
\tau_c(t) =  \frac{\delta t}{4H+2}\left( \frac{t}{\delta t} \right)^{2H+1}
h_{H+1/2}(x)\ ; \quad x= \delta t / t\ ,
\ee
and, for $x < 1$, $h_{H+1/2}(x)$ is again given by Eq. ~\eqref{aux_function}, but
with $H+1/2$ instead of $H$.
Then, an asymptotic formula similar to Eq. ~\eqref{aux_func_taylor} can be
derived:
\be
h_{H+1/2}(x) = 2H (2H+1) x^2 + O(x^4)\ ,
\ee
and, finally:
\be
\tau_c(t) = H (\delta t)^{2-2H} t^{2H-1}\ \ \mathrm{for} \ \ t \rightarrow \infty
\ .
\label{asympt_tauc}
\ee
Clearly, the mathematical limit $t \rightarrow \infty$ corresponds to the
physical regime $t \gg \delta t$.
Exploiting the asymptotic behavior of $\tau_c(t)$ given in Eq.
~\eqref{asympt_tauc}, we can now derive the values of the correlation time
scale $\tau_c=\tau_c(\infty)$:
\be
\tau_c=\lim_{t\rightarrow\infty}\tau_c(t) = \left\{
\begin{array}{l}
  + \infty\ ; \quad \quad\  1/2 < H \le 1
  \ ;\\
  \ \\
  \delta t / 2 < \infty\ ; \quad \quad  H = 1/2
  \ ;\\
\ \\
0\ ; \quad \quad  0 < H < 1/2
\ .
\end{array}
\right.
\label{tauc_estimate}
\ee
The three cases correspond to persistent (superdiffusive) FBM,
normal Brownian motion and antipersistent (subdiffusive) FBM, respectively.


\noindent\fbox{%
    \parbox{\textwidth}{
    \small
 {\bf Box 2. Properties of $R(t)$ and $g(\tau)$}
 
\noindent
 The use of Laplace transform, defined by the expression: 
\be
\widetilde{u}(s) = {\cal L}_{t \rightarrow s}[u(t)](s) = 
\int_0^\infty e^{-st}u(t)dt\,,
\nonumber
\ee
gives important information about the normalization and moments of distributions.
The stationary correlation function $R(t)$ and the distribution $g(\tau)$ are related
by Eq. (\ref{superpos_corr_eq}). 
For any choice of the distribution $g(\tau)$, the correlation function $R(t)$ and $g(\tau)$ must satisfy the following properties:
\begin{itemize}
\item[(i)]
  The distribution $g(\tau)$ must be a PDF normalized to $1$:
\begin{equation}
\widetilde{g}(0)=1\nonumber\,,\nonumber
\label{lapl_norm}
\end{equation}
which determines a constrain on the
behavior of the first derivative of the correlation function:
%
\begin{equation}
\lim_{s \to +\infty} s \cdot {\cal L}\left[ -\frac{dR(t)}{dt} \right](s)= 
-\frac{dR}{dt}(0_+) = \la \nu \ra  \, .\nonumber
\label{lapl_norm_2}
\end{equation}
\item[(ii)]
  The MSD is a power-law of time
with superdiffusive scaling $1< \phi < 2$ in the asymptotic long-time limit:
\begin{equation}
\lim_{t \to \infty} \frac{\sigma^2_X(t)}{t^\phi} = C_1\ ;\quad 
\lim_{s \to 0} s^{1+\phi} \cdot \widetilde{\sigma^2_X}(s) = C_2 
\ .
\label{lapl_anom_diff_1}
\end{equation}
where $C_1$ and $C_2$ are proper constants and the second asymptotic limit follows
from the Tauberian theorem \cite{feller-1971}.
From Eq. (\ref{taylor21_1}) or Eq. (\ref{taylor21_2}) it results:
\begin{equation}
\frac{d^2\sigma_X^2(t)}{dt^2} = 2\, R(t) \ ;\quad
\widetilde{\sigma^2_X}(s) = \frac{2}{s^2} \widetilde{R}(s) \ ,\nonumber
\label{lapl_var_corrv}
\end{equation}
we get equivalently the following expression for the stationary correlation 
function:
\begin{equation}
\lim_{t \to \infty} \frac{R(t)}{t^{\phi-2}} = C_3 \ ;\quad
\lim_{s \to 0}  s^{1-\eta} \cdot \widetilde{R}(s) = C_4
\label{lapl_anom_diff_2}
\end{equation}
with $\eta =2-\phi$, $0<\eta<1$.
Note that the above limits can be equivalently written as asymptotic behaviors,
e.g.: $R(t) \sim t^{\phi-2}$ for $t\rightarrow \infty$, which means that the
function $R(t)$ is approximated by $C_3 t^{\phi-2}$ in the long time range.
\item[(iii)]
  The MSD at time zero is zero:
\begin{equation}
\lim_{t \to \infty} \sigma^2_X(t) = 0\, ;\quad \lim_{s \to +\infty}  s \cdot 
\widetilde{\sigma^2_X}(s) = 0
\nonumber
\end{equation}
\item[(iv)]
  Furthermore being $0 < R(0) < \infty$, from Eq.(\ref{var_stat_av})
  the distribution $g(\tau)$ must have non-zero, finite mean:
\begin{equation}
\lim_{s \to +\infty} s \cdot \widetilde{R}(s) = R(0) \propto \langle \tau \rangle\,.\nonumber
\label{lapl_finite_tau}
\end{equation}
\end{itemize}
    }
}

\newpage
\section{Derivation of the PDF $g(\tau)$}
\label{appendix_gtau}

  The properties that must be satisfied by the stationary correlation function
  $R(t)$ and by the PDF $g(\tau)$ are listed in the above Box 1.

\vspace{.4cm}
\noindent
We now prove the following

\vspace{.3cm}
\noindent
{\bf Theorem (PDF $g(\tau)$)}

\noindent
Given Eq. (\ref{superpos_corr_eq}) defining the stationary correlation function
of the Langevin equation with random parameters, Eq. (\ref{lang_free}),
the PDF $g(\tau)$ given in Eq. (\ref{pdf_gtau}) satisfies all the required
constrains (i-iv) listed in Box 1.

\vspace{.3cm}
\noindent
{\bf Proof:}

\vspace{.3cm}
\noindent
{\bf (i) normalization and (iv) finite mean}:

Let us write:
$$
g(\tau) = \frac{C}{\tau} L_\eta^{-\eta} \left( \frac{\tau}{\tau_*} \right)\ ,
$$
where $\tau_*$ must be introduced to get an adimensional parameter as argument
of $L_\eta^{-\eta}$.
The mean correlation time is given by:
\begin{equation}
\la \tau\ra = \int_0^\infty \tau g(\tau)d\tau =
 C \int_0^\infty L_\eta^{-\eta}\lf( \frac{\tau}{\tau_*} \rg) d\tau 
= C \tau_*\ ,
\end{equation}
so that we have:
\be
g(\tau) = \frac{C}{\tau} L_\eta^{-\eta} \left( C \frac{\tau}{\la \tau \ra} 
\right)\ .
\label{ansatz}
\ee
The normalization constant $C$ can be obtained by imposing 
${\cal L}[g(\tau)](0)=1$. Exploiting the relationship
$\int_s^\infty \exp(-\xi \tau) d\xi = \exp(-\xi\tau)/\tau$ and making the
change of variables $\tau = (\la\tau\ra / C) \tau^\prime$, we get:
\begin{equation}
\begin{split}
{\cal L}[g(\tau)](s) &=C \cdot \int_{s\la\tau\ra/C}^\infty  {\cal L}[L_\eta^{-\eta}(\tau)](\xi)d\xi \\
  & = C\cdot\int_{s\la\tau\ra/C}^\infty e^{-\xi^\eta}d\xi = \\
  & {(\tiny  x=\xi^\eta)}\\
  & = C \frac{1}{\eta}\, \int_{s\la\tau\ra/C}^\infty \frac{1}{\eta}e^{-x}x^{1/\eta -1}
dx\ ,
\end{split}
\end{equation}
and:
\begin{equation}
\begin{split}
{\cal L}[g(\tau)](0) &=C\cdot\int_0^\infty \frac{1}{\eta}e^{-x}x^{1/\eta -1}dx\\
&=C\cdot\frac{\Gamma(1/\eta)}{\eta}  = 1\ .
\end{split}
\end{equation}
Substituting this relationship into Eq. (\ref{ansatz}) we finally get
Eq. (\ref{pdf_gtau}), which is a properly normalized PDF.

\vspace{.3cm}
\noindent
{\bf (ii) superdiffusive scaling}:

We now prove that $R(t)\sim t^{-\eta}$, with $0<\eta<1$, a condition leading to 
the superdiffusive scaling for the position variance: 
$\sigma_X^2(t)\sim t^\phi$, $1<\phi = 2 -\eta <2$.
This can be proven thanks to the integral
representation of the extremal L\'evy density:
\begin{equation}
 L_\eta^{-\eta}(x)=\frac{1}{\eta x}\frac{1}{2\pi
   i}\int_{\gamma-i\infty}^{\gamma+i\infty}
 \frac{\Gamma(s/\eta)}{\Gamma(s)}x^sds ,\quad 0<\eta<1\ .
\end{equation}
Hence, we have:
\begin{equation}
\begin{split}
R(t)&=\la \nu \ra \frac{\eta}{\Gamma(1/\eta)}\int_0^\infty e^{-t/\tau}
L_\eta^{-\eta}\lf(\frac{\tau}{\tau_*}\rg) d\tau
\\ 
&=\la \nu \ra\frac{\eta}{\Gamma(1/\eta)} \int_0^\infty e^{-t/\tau}\lf[
  \frac{1}{\eta}\frac{1}{2\pi i}\int_{\gamma-i\infty}^{\gamma+i\infty}
  \frac{\Gamma(s/\eta)}{\Gamma(s)}\lf(\frac{\tau}{\tau_*}\rg)
  ^{(s-1)}ds\rg]d\tau
\\ 
&=\la \nu \ra \frac{\eta}{\Gamma(1/\eta)}\frac{1}{\eta }\frac{1}{2\pi
  i}\int_{\gamma-i\infty}^{\gamma+i\infty} \frac{\Gamma(s/\eta)}{\Gamma(s)}
\lf[\int_0^\infty e^{-t/\tau}\lf(\frac{\tau}{\tau_*}\rg)^{s-1}d\tau\rg]ds = 
\\ 
&({\tiny  \xi = t/\tau})
\\ 
&=\la \nu \ra \la \tau\ra \frac{1}{\eta
}\frac{1}{2\pi i}\int_{\gamma-i\infty}^{\gamma+i\infty}
\frac{\Gamma(s/\eta)}{\Gamma(s)}\lf[\int_0^\infty e^{-\xi}\xi^{-1-s}
  \lf(\frac{t}{\tau_*}\rg)^s d\xi\rg]ds
\\ 
&=\la \nu \ra \la \tau\ra
\frac{1}{\eta }\frac{1}{2\pi i}\int_{\gamma-i\infty}^{\gamma+i\infty}
\frac{\Gamma(s/\eta)\Gamma(-s)}{\Gamma(s)} \lf(\frac{t}{\tau_*}\rg)^s ds\ ,
\end{split}
\end{equation}
where $\tau_* =  \la\tau\ra\, \Gamma(1/\eta)/\eta$.
It is useful to rewrite the expression as:
\begin{equation}
\begin{split}
 R(t)&=\la \nu \ra \la \tau\ra \frac{1}{\eta }\frac{1}{2\pi
   i}\int_{\gamma-i\infty}^{\gamma+i\infty}
 \frac{(\eta/s)\Gamma(s/\eta+1)\Gamma(-s)}{(1/s)\Gamma(s+1)}
 \lf(\frac{t}{\tau_*}\rg)^s ds\\ &=\la \nu \ra \la \tau\ra
 \frac{1}{2\pi i}\int_{\gamma-i\infty}^{\gamma+i\infty}
 \frac{\Gamma(s/\eta+1)\Gamma(-s)}{\Gamma(s+1)}
 \lf(\frac{t}{\tau_*}\rg)^s ds\ ,
\end{split}
\end{equation}
which can be solved through the residues theorem considering the poles
$s/\eta+1=-n$ or $s=n$, with $n=0,1,2..\infty$.

In the first case we have:
\begin{equation}
\begin{split}
 R(t)=&\la \nu \ra \la \tau\ra \sum_{n=0}^\infty
 \eta\frac{(-1)^n}{n!}\frac{\Gamma(\eta (n+1))}{\Gamma(1-\eta
   (n+1))}\lf(\frac{t}{\tau_*}\rg)^{-\eta (n+1)}\\ =&\la \nu \ra \la
 \tau\ra \sum_{n=1}^\infty \frac{(-1)^n}{n!}\frac{\Gamma(\eta
   n)}{\Gamma(-\eta n)}\lf(\frac{t}{\tau_*}\rg)^{-\eta n}
\end{split}
 \end{equation}
where each term of the series is obtained by the limit:
\begin{equation}
\begin{split}
&\lim_{s \rightarrow -\eta(n+1)}(s+\eta(n+1))\frac{\Gamma(s/\eta+1)\Gamma(-s)}
{\Gamma(s+1)}\lf(\frac{t}{\tau_*}\rg)^s
\\ 
&\lim_{s \rightarrow -\eta(n+1)}\eta((s/\eta+1)+n)\frac{\Gamma(s/\eta+1)\Gamma(-s)}
{\Gamma(s+1)}\lf(\frac{t}{\tau_*}\rg)^s
\\ 
&\lim_{s \rightarrow -\eta(n+1)}\frac{\eta((s/\eta+1)+n)}{(s/\eta+1)_{n+1}}
\frac{\Gamma(s/\eta+n+2)\Gamma(-s)}{\Gamma(s+1)}\lf(\frac{t}{\tau_*}\rg)^s
\\ 
&\lim_{s \rightarrow - \eta(n+1)}\frac{\eta(-1)^n}{n!}\frac{\Gamma(\eta(n+1))}
{\Gamma(1-\eta(n+1))}\lf(\frac{t}{\tau_*}\rg)^{-\eta (n+1)}
\end{split}
\end{equation}
When $t\rightarrow\infty$ only the first term survives and we find:
\begin{equation}
 R(t)=\la \nu \ra \la \tau\ra \frac{\Gamma(\eta+1)}{\Gamma(1-\eta
   )}\lf(\frac{t}{\tau_*}\rg)^{-\eta}\ .
\end{equation}
Substituting $\tau_* =  \la\tau\ra\, \Gamma(1/\eta)/\eta$, we finally get
Eq. (\ref{corr_free}), from which we obtain the superdiffusive scaling of 
the position variance $\sigma^2_X(t)\propto t^\phi$, with $\phi=2-\eta$.

Considering the poles in the other semi-plane, $s=n$ with
$n=0,1,2..\infty$, we find that:
\begin{equation}
 R(t)=\la \nu \ra \la \tau\ra \frac{1}{\eta}\sum_{n=0}^\infty
 \frac{(-1)^n}{n!}\frac{\Gamma(n/\eta)}{\Gamma(n)}\lf(\frac{t}{\tau_*}\rg)^{n}
\end{equation}
converges to $R(0)=\la \nu \ra \la \tau\ra$, as already shown before.

\vspace{.3cm}
\noindent
{\bf (iii) MSD at time zero is zero}:

The condition  $\sigma^2_X(t = 0) = 0$ is clearly verified.

\vspace{.5cm}
\noindent
{\bf Example:}\\
In the special case $\eta=1/2$, the extremal L\'evy function corresponds
to the L\'evy--Smirnov distribution, the whole exercise can be solved
analitycally and we may consider for simplicity $\la
\tau\ra\frac{\Gamma(1/\eta)}{\eta}=1$ :
\begin{equation}
 g(\tau)=\frac{1}{\sqrt{4\pi\tau^5}}e^{-1/(4\tau)}
\end{equation}
Solving the integral the analytical form of the correlation function
turns to be:
\begin{equation}
 R(t)=\frac{\Gamma(1/2)}{\sqrt{4\pi}}\lf(t+\frac{1}{4}\rg)^{-1/2}
\end{equation}
which leads to the following exact formula for the position variance:
\begin{equation}
 \sigma_X^2(t)=\frac{\Gamma(1/2)}{\sqrt{\pi}}\lf[\frac{4}{3}\lf(t+\frac{1}{4}\rg)^{3/2}-t-\frac{1}{6}\rg]\ ,
\end{equation}
satisfying both superdiffusive long-time scaling and $\sigma_X^2(0)=0$ 
conditions.

\vspace{.3cm}
{\bf NOTE:} {\it The Einstein--Smoluchovsky relation}

By substituting Eq. (\ref{corr_v_stat}) into Eq. (\ref{tau_corr}) it is
easy to see that $\tau_c=\tau$.
Using the following equation (see the last equation in Box 1 of the Main Text):
\be
R(0 | V_0,\tau,\nu) = \langle V^2 | V_0,\tau,\nu \rangle_{\rm st} = \nu \tau\ ,
\nonumber
\ee
and substituting Eq. (\ref{corr_v_stat}) into 
the definition of $D_{_X}$, Eq. (\ref{norm_diff_2}),
we get the Einstein--Smoluchowsky relation:
\be
D_{_X} = \nu \tau^2 = \tau\  \la V^2 | V_0,\tau,\nu \ra_{\rm st}\ ,
\label{einstein_smo}
\ee
which, apart from the conditional statistics, is essentially the same as
Eq. (\ref{einstein_smo_taylor}).
For a standard OU process with fixed $\nu$ and $\tau$, 
$\langle V^2 | V_0,\tau,\nu \rangle_{\rm st}=\langle V^2 \rangle_{\rm eq}$ and
Eq. (\ref{einstein_smo}) relates the diffusion ($D_{_X}$) and relaxation 
($\tau$) properties through the equilibrium distribution 
($\langle V^2 \rangle_{\rm eq}$).
In his 1905 paper \cite{einstein_adp05}, Einstein studied the Brownian motion 
in a gas at equilibrium, where velocity distribution is given by the 
Maxwell--Boltzmann law. In this case, the Einstein--Smoluchowsky relation
becomes:
\be
D_{_X} = \tau\ \langle V^2 \rangle_{\rm st} = \tau\ \frac{k T}{m} \ ,
\label{einstein_smo_mb}
\ee
being $T$, $m$ and $k$ the gas temperature, the Brownian particle mass and
the Boltzmann constant, respectively.

\section{Numerical scheme for the Langevin equation}
\label{num_scheme}

\noindent
In order to avoid stability problems, the numerical algorithm for the 
simulation of Eqs. (\ref{diff_eq}) and \ref{lang_free}) was implemented 
using an implicit scheme with order of strong convergence $1.5$ 
\cite{kloeden_platen-1992}. This is given by the following expression:
\begin{eqnarray}
V_{n+1} = V_n + b \Delta W_n &+& \frac{1}{2} \{a(V_{n+1}) + a (V_n) \}+
\label{num_scheme_1}
\\
&+& 
\frac{1}{2\sqrt{\Delta t}} \{a(\overline{V}_+) - a(\overline{V}_-) \} 
\left( \Delta Z_n - \frac{1}{2} \Delta W_n \Delta t \right) \, ,
\nonumber
\end{eqnarray}
being $V_n=V(n \Delta t)$, $\Delta t$ the time step, 
$\Delta W_n = W(t_n+\Delta t) - W(t_n)$ the increments of the Wiener process,
$a(V) = -V/\tau$ and $b=\sqrt{2\nu}$ the drift and noise terms, respectively.
Further, we have:
\begin{equation}
\begin{split}
&\overline{V}_{\pm}=V_n+a(V_n) \Delta t \pm b \sqrt{\Delta t} \, , \\
&\Delta Z_n = \frac{1}{2} (\Delta t)^{3/2} \left( u_1(n) + \frac{1}{\sqrt{3}} 
u_2(n) \right) \, ,
\end{split}
\label{num_scheme_2}
\end{equation}
being $u_1(n)$ and $u_2(n)$ two independent random numbers with uniform 
distributions in $[0,1]$.
A suitable time step $\Delta t$, also depending on the time scale $\tau$, 
is necessary to maintain the accuracy of the numerical scheme.
To take into account both the ensemble variability of the relaxation time 
$\tau$, which is different for different trajectories, and the time variability 
of drift and noise terms along the
same trajectory, we applied a variable time step according to the scheme given
in Ref. \cite{thomson-jfm-1987}:
\begin{equation}
\Delta t = \min \left\{\frac{0.05}{b}, \frac{0.1}{|a|} \right\} \, .
\label{time_step}
\end{equation}
This adaptive time step allows to avoid any problem of convergence and accuracy
in the numerical scheme, Eqs. (\ref{num_scheme_1}) and (\ref{num_scheme_2}).
At the same time, in the range of short $\tau$, this algorithm can give very
short time steps, thus determining very long simulation times for
a consistent number of trajectories.
To overcome this problem we note that the short time regime $\tau\ll\la\tau\ra$
of the PDF $g(\tau)$ does not significantly affect the anomalous 
scaling of diffusion, which mostly depends on the asymptotic tail of the 
distribution $g(\tau)$. A cut-off was then introduced in the short-time regime. 
By comparing the numerical simulations with theoretical results we 
chose the cut-off value $\tau_{\rm min}=0.004$, much smaller that $\la\tau\ra$,
which is always of the order $0.5-1$ for all sampled sets of $\tau$.

\section{Numerical algorithm for the random generator of $\tau$}
\label{appendix_num}


\noindent
Here we describe a method to generate random variables $\tau$ 
distributed according to the law of Eq. (\ref{pdf_gtau}),
\begin{equation}
\label{eq_tarpdf}
g(\tau) = A(\eta) L_\eta^{-\eta}(\tau)/\tau,
\end{equation}
where $A(\eta)$ is the normalization coefficient, and $\tau$ is
already dimensionless.

For this, we use a well-known inverse transform sampling method (see,
e.g.~\cite{devroye2006nonuniform}), so the procedure is
straightforward.

\begin{figure}
\includegraphics[width=0.49\textwidth, height=0.24\textheight]{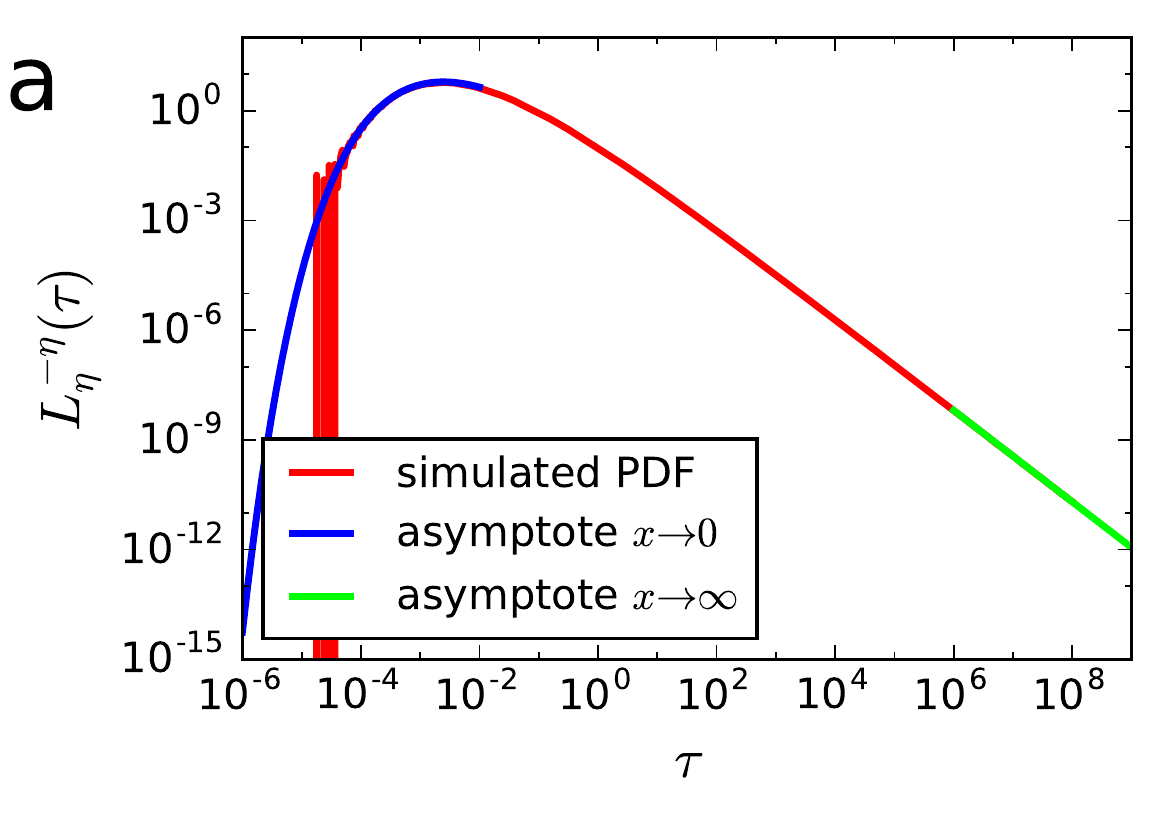}
\includegraphics[width=0.49\textwidth, height=0.24\textheight]{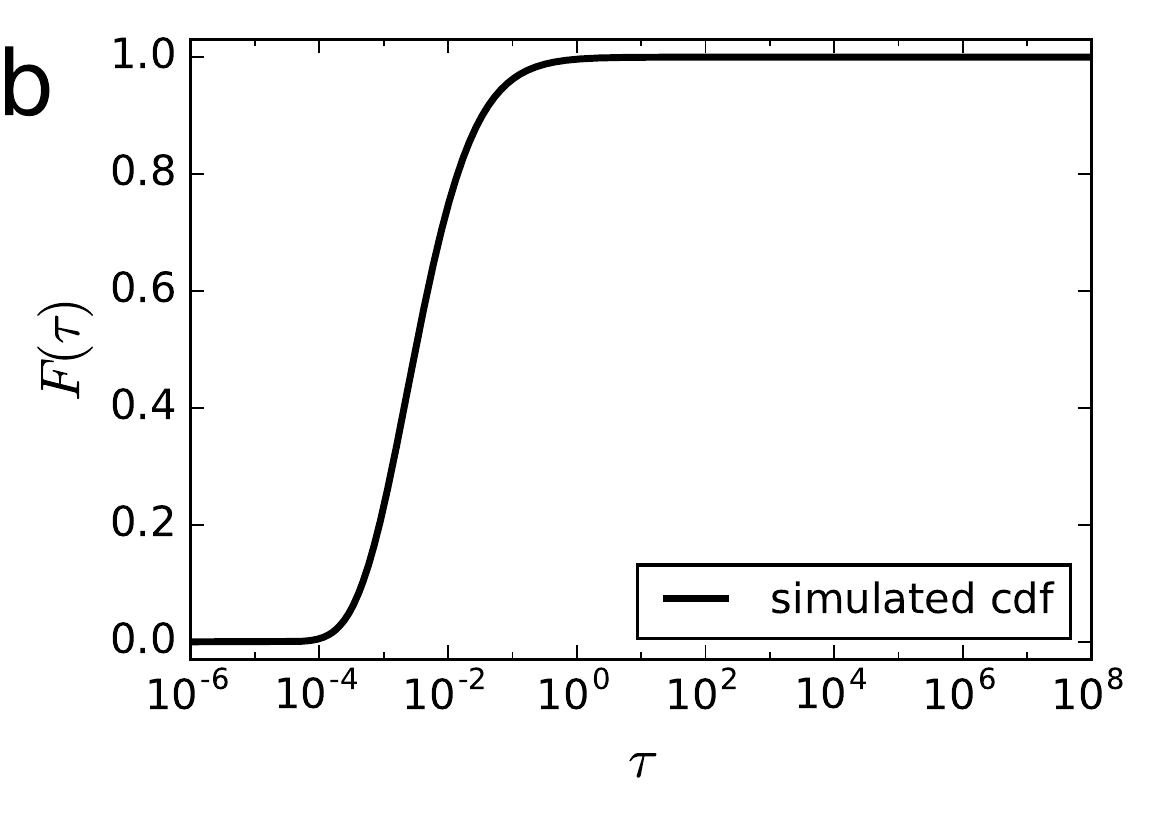}
\caption{\label{fig_pdfs}(color online) (a) Simulated L\'evy extremal
  density (red) together with asymptotics at small arguments (blue)
  and large ones (green). (b) Cumulative distribution function of
  $g(\tau)$.}
\end{figure}

First, we generate a set of extremal L\'evy density random numbers
$L_\eta^{-\eta}(\tau)$ by using the generator described in
Refs.~\cite{chambers_etal-jasa-1976, weron-spl-1996}, see Eq.~(3.2) of
the latter paper, and extract its histogram. Since the beginning of
the histogram has much statistical noise (red curve in
Fig.~\ref{fig_pdfs}a), it is a good solution to replace these values
with analytical asymptote at small arguments
\cite{mainardi_etal-fcaa-2001} (blue curve in
Fig.~\ref{fig_pdfs}a). Moreover, we also expand the histogram with
another asymptote, at large $\tau$s (green curve in
Fig.~\ref{fig_pdfs}a):
\begin{eqnarray}
L_\eta^{-\eta} (\tau) &\sim & A_1 \tau^{-a_1} \exp(-b_1 \tau^{c_1}),
\qquad \tau\to 0^+,\\
\label{asy_infty} L_\eta^{-\eta}(\tau) &\sim & \frac{C_1(\eta)}{|\tau|^{1+\eta}}, \qquad \tau \to \infty,
\end{eqnarray}
where
\begin{eqnarray}
&&A_1 = \left\{\left[2\pi(1-\eta)\right]^{-1}
  \eta^{1/(1-\eta)}\right\}^{1/2},\\ &&a_1 = \frac{2-\eta}{2(1-\eta)},
  \quad b_1=(1-\eta)\eta^{\eta/(1-\eta)}, \quad
  c_1=\frac{\eta}{1-\eta}\\ &&\label{asy_infty_coef} C_1(\eta) \approx
  \frac{1}{\pi} \sin \left(\pi \frac{\eta}{2}\right) \Gamma(1+\eta).
\end{eqnarray}

Then, we divide the obtained histogram by argument and find the
normalization coefficient numerically in order that the resulting PDF
is normalized to unity. Finally, we calculate the semi-analytical
cumulative distribution function (CDF) (see Fig.~\ref{fig_pdfs}):
\begin{eqnarray}
F(\tau_k) &=& \sum \limits_{i=0}^k g(\tau_i) \delta \tau_i, \qquad
\tau_k \leq \tau_n;\\ F(\tau) &=& \sum \limits_{i=0}^n g(\tau_i)
\delta \tau_i + \int \limits_{\tau_n}^\tau
\frac{A(\eta)}{\tau'^{2+\eta}} d\tau', \qquad \tau > \tau_n,
\end{eqnarray}
where $\delta \tau_i$ is the $i^\mathrm{th}$ histogram's bin width, $i
= 0, 1, 2.. n$.

Now, we draw a random variable $\tau$ obeying the target pdf~
\eqref{eq_tarpdf} with
\begin{equation}
\tau = F^{-1}(u),
\end{equation}
where $u\in [0,1)$ is a uniformly distributed random variable: $F^{-1}$ is a 
numerically (or if $u>F(x_n)$, semi-analytically) inverted CDF.

\begin{figure}
\includegraphics[width=0.48\textwidth, height=0.24\textheight]{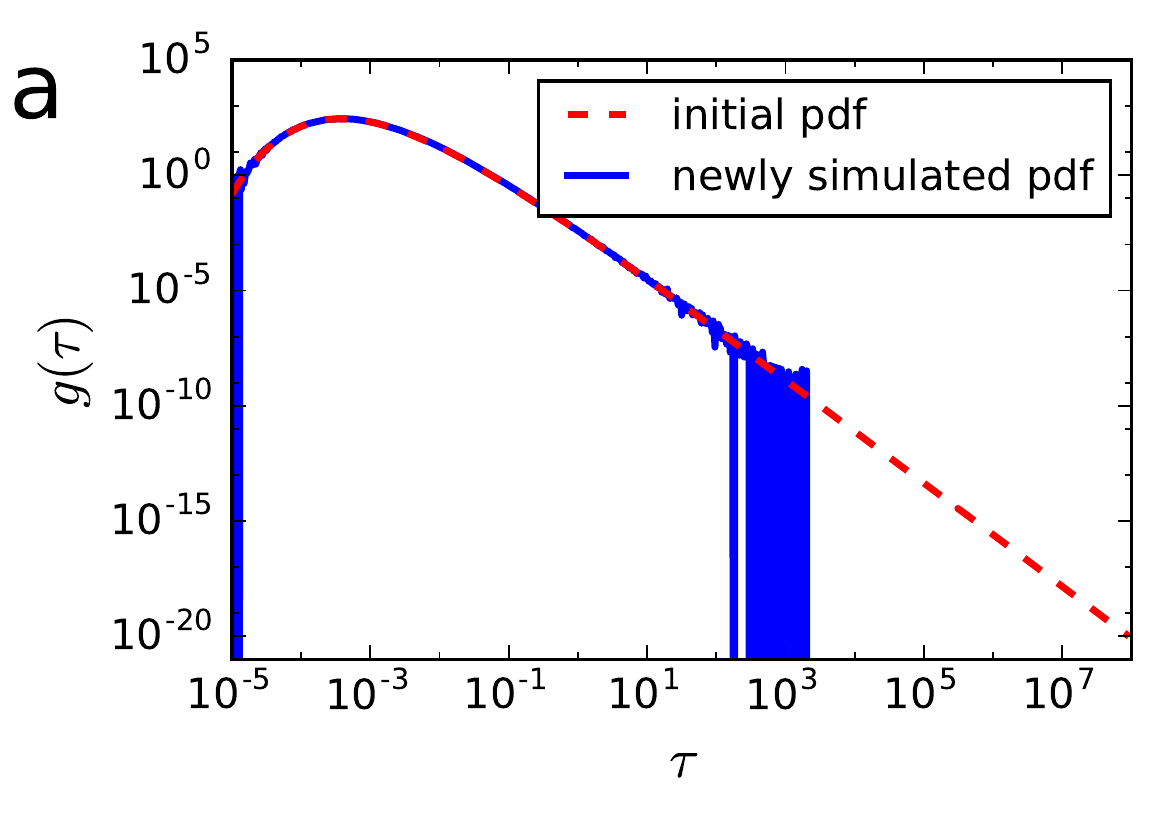}
\includegraphics[width=0.48\textwidth, height=0.24\textheight]{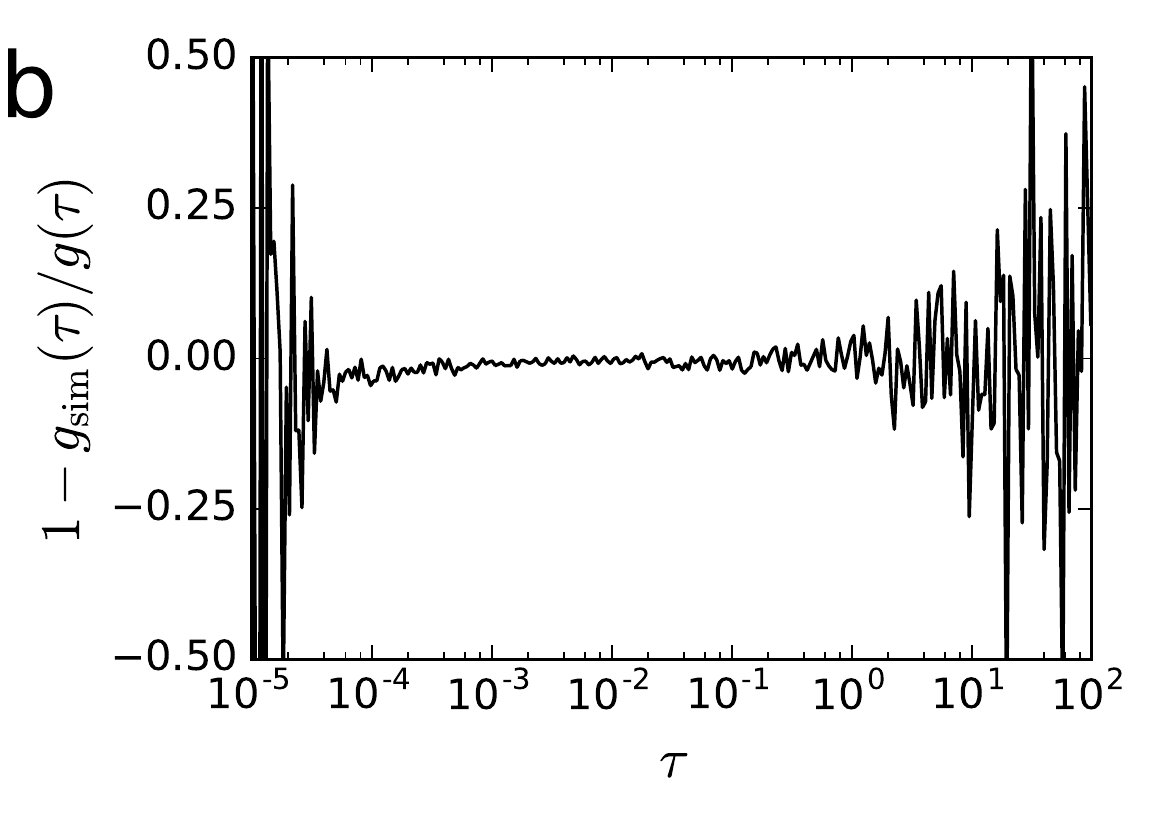}
\caption{\label{fig_pdfsres}(color online) (a) Comparison of the original PDF~\eqref{eq_tarpdf} (red) and the PDF histogram of generated numbers (blue). (b) Relative error between original and simulated PDFs.}
\end{figure}

Let us take out a verification and compare the original PDF $g(\tau)$
used for the simulations and the histogram of the generated $10^7$
random numbers with this algorithm $g_\mathrm{sim}(\tau)$. The result
is shown in Fig.~\ref{fig_pdfsres}. At intermediate values of $\tau$
the inaccuracy is about $1\%$, increasing due to statistical error at
very small and large $\tau$s (where $g(\tau)$ is small).

\vspace{.5cm}
\noindent
      The software for the numerical simulations were written in C++ language
      (Debian gcc 4.9) and Python 2.7 and can be downloaded at the following
      web-site:
      { https://gitlab.bcamath.org/opensource/lecm}.
      
      The codes include the algorithms described in this section and in the
      previous one.
      The simulation runs were performed on  computational facilities of
BCAM-Basque Center for Applied Mathematics.

\section{Schneider grey noise, gBM and ggBM}
\label{grey_noise}

We here provide an intuitive presentation of the Schneider grey noise,
the grey Brownian motion and the generalized grey Brownian motion. 
More rigouros details can be found in 
\cite{schneider_1990,schneider_1992,mura-phd-2008,mura_etal-jpa-2008,mura_etal-pa-2008,mura_etal-itsf-2009,grothaus_etal-jfa-2015,grothaus_etal-jfa-2016}.
 
The grey noise is a generalization 
on the basis of the Mittag--Leffler function 
of the white noise.
The Mittag-Leffler function $E_\beta(z)$ is defined as
\be
E_\beta(z)=\sum_{n=0}^\infty \frac{z^n}{\Gamma(\beta n +1)} \,,
\ee 
and it is a generalization of the exponential function 
that is recovered as special case when $\beta=1$,
i.e., $E_1(-z)=\e^{-z}$.
As well as the exponential function,
when $0 < \beta < 1$, the Mittag--Leffler function is a completely monotonic function. 
A useful formula for what follows is 
\be
\left.
-\frac{d^2}{dz^2} E_\beta(-z^2 \, q) \right|_{z=0}= \frac{2}{\Gamma(1+\beta)} \, q \,.
\label{formulaML}
\ee

For any characteristic functional
$\Phi(z)$ there exists a unique probability measure $\mu$ such that
\be
\Phi(z)=\int_{-\infty}^{+\infty} \e^{i z \tau} \, d\mu(\tau) \,,
\ee
and if $\Phi(z)=E_\beta(-z^2)$, $0 < \beta < 1$,
the probability measure $\mu$ is the so-called Schneider grey noise 
\cite{schneider_1990,schneider_1992,mura_etal-itsf-2009}.
When $\beta=1$ we have $E_1(-z^2)=\e^{-z^2}$,
and the Gaussian white noise follows. 

Let us introduce the stochastic process $X(t)$ driven by the noise $\mu$
and we look for its probability density function.
The characteristic function is
\be
\langle \e^{i z X(t)} \rangle =
\int_{-\infty}^{+\infty} \e^{i z X(t)} \, d\mu(t) = E_\beta (-z^2 \, \varphi^2_\alpha(t)) \,,
\label{FT}
\ee
where function $\varphi_\alpha(t)$ takes into account what remains of parameter $t$
after the integration, and it is related to the scaling in time of $X(t)$.
By the inversion of (\ref{FT}) we have the probability density function of $X(t)$
as follows
\be
p(x,t)=\frac{1}{2\pi}
\int_{-\infty}^{+\infty} \e^{-i z x} E_\beta (-z^2 \, \varphi_\alpha^2(t)) dz =
\frac{1}{2 \, \varphi_\alpha(t)} 
M_{\beta/2}\left(\frac{|x|}{\varphi_\alpha(t)}\right) \,,
\ee
where $M_{\beta/2}$ is the M-Wrigth/Mainardi function.
By using (\ref{FT}) and (\ref{formulaML}), we have that the variance of $X(t)$ is 
\be
\langle x^2 \rangle = 
\left.
-\frac{d^2}{dz^2} E_\beta(- z^2 \, \varphi^2_\alpha(t)) \right|_{z=0}
=\frac{2}{\Gamma(1+\beta)} \varphi^2_\alpha(t) \,.
\label{varianza}
\ee

In the same spirit, the correlation function of the process $X(t)$ can be computed.
In fact from (\ref{FT}) it holds
\be
\langle \e^{i z [X(t)-X(s)]} \rangle =
\int_{-\infty}^{+\infty} \e^{i z [X(t)-X(s)]} \, d\mu(t,s) = 
E_\beta (-z^2 \, \varphi_\alpha^2(t,s)) \,,
\label{correlation}
\ee
and 
by applying again formula (\ref{formulaML}) 
the correlation function results to be 
\be
\frac{1}{\Gamma(1+\beta)} \, 
(\varphi_\alpha(t) + \varphi_\alpha(s) - \varphi_\alpha(t,s)) \,. 
\label{correlazione}
\ee

Now we discuss how to establish function $\varphi_\alpha(t)$.
Let $\mathbbm{1}_{[a,b]}$ be the indicator function 
such that it is equal to $1$ when $a<t<b$ and to $0$ elsewhere. 
In analogy with the Wiener process where the Brownian motion
is $\displaystyle{\mathcal{B}(t)=\int_0^t dW(\tau)}$, 
we write the process $X(t)$ as 
\be
X(t)=\int_0^t d\mu(\tau)=\mathbbm{1}_{[0,t]} \, X_0(\mathbbm{1}_{[0,t]}) \,,
\label{prodotto}
\ee 
where $X_0$ is a random variable equivalent in distribution to $X(t)$
but independent of $t$, i.e., the probability density function of $X_0$
is $p_0(x)=p(x,t=1)$.
From (\ref{prodotto}) we have that 
\be
\langle [X(t)]^2 \rangle=
\langle [\mathbbm{1}_{[0,t]}]^2 \rangle \, \langle [X_0]^2 \rangle \,,
\ee
and from comparison with (\ref{varianza}) and (\ref{correlazione}), 
we obtain that $\varphi_\alpha(t)$
is established through the stochastic process $\mathbbm{1}_{[a,b]}$ that meets
\be
\langle [\mathbbm{1}_{[0,t]}]^2 \rangle = 
\varphi_\alpha^2(t) \,, 
\ee
\be
\langle \mathbbm{1}_{[s,t]}\mathbbm{1}_{[0,s]} \rangle = 
\frac{1}{2} (\varphi_\alpha^2(t) + \varphi_\alpha^2(s) 
- \varphi_\alpha^2(t,s)) \,.
\ee

Finally we observe that, by setting $\varphi^2_\alpha(t) = t^\alpha$,
$X(t)$ is the Brownian motion when $\alpha=\beta=1$,
and we refer to it as the grey Brownian motion and the generalized grey Brownian motion 
when $0< \alpha=\beta < 1$ and  
$0 < \alpha < 2$, $0 < \beta < 1$, respectively.
Moreover, 
in order to have a process with stationary increments we assume 
$\varphi^2_\alpha(t,s)=|t-s|^\alpha$, 
the correlation function results to be 
\be
\frac{1}{\Gamma(1+\beta)} \, (t^\alpha + s^\alpha - |t-s|^\alpha) \,. 
\ee

The corresponding stochastic process is obtained with 
a randomly-scaled Gaussian process,
i.e., a Gaussian process multiplied for 
a non-negative independent randon variable not dependent on time. 

From integral representation formulae of the M function 
\cite{mainardi_etal-fcaa-2003},
we have that $X_0$ has the same density of $X(1)$ if, for example, 
we state $X_0=\sqrt{\Lambda} \, \mathcal{B}(1)$ where $\Lambda$ is a non-negative
random variable distributed according to $M_\beta$ 
and $\mathcal{B}(1)$ is a Gaussian variable. 
Finally, we obtain that
\be
X(t)= \sqrt{\Lambda} \,\, \mathbbm{1}_{[0,t]} \, \mathcal{B}(\mathbbm{1}_{[0,t]}) \,.
\ee
Looking at (\ref{varianza}) and (\ref{correlazione}), the process
$\displaystyle{\mathbbm{1}_{[0,t]} \, \mathcal{B}(\mathbbm{1}_{[0,t]})}$ 
is the fractional Brownian motion $X_H(t)$ \cite{biagini_etal-2008}
characterized by
\be
\langle [X_H(t)]^2 \rangle = 
t^{2H} \,, 
\ee
\be
\langle X(t)X(s) \rangle = 
\frac{1}{2} (t^{2H} + s^{2H} - |t-s|^{2H}) \,.
\ee
Finally, by setting $H=\alpha/2$, 
the trajectories of the process $X(t)$ can be generated by
\be
X(t) = \sqrt{\Lambda} \, X_H(t) \,.
\ee
Since the fBm $X_H(t)$ is fully characterized by the variance and the correlation functio,
the process $X(t)$ is also fully characterized by the variance and the correlation function.

\noindent
With a somewhat forced terminology, the term ggBM can be thought to include any
randomly scaled Gaussian process, i.e.,
any processes defined by the product of a Gaussian process 
with an independent and constant non-negative random variable.

\section{Mainardi distribution and L\'evy densities}
\label{mainardi}

\noindent
Fractional diffusion processes are a generalization of classical Gaussian diffusion,
mainly in the direction of the time-fractional diffusion, i.e.,
by replacing the first derivative in time with a time-fractional derivative,
and in the direction of the space-fractional diffusion,
i.e., by replacing the second derivative in space with a space-fractional derivative.
In the case of time-fractional diffusion the Gaussian particle density is generalized by
the so-called $M$-Wright/Mainardi functions \cite{mainardi_etal-ijde-2010,pagnini-fcaa-2013},
and in the case of the space-fractional diffusion the particle density
is generalized by the so-called L\'evy stable densities \cite{mainardi_etal-fcaa-2001}.

\noindent
The M-Wright/Mainardi function $M_\nu(r)$, $r \ge 0$, $0< \nu < 1$,
is defined by the series:
\be
M_\nu(r)=\sum_{n=0}^\infty \frac{(-r)^n}{n!\Gamma[-\nu n + (1-\nu)]}
=\frac{1}{\pi} \sum_{n=0}^\infty \frac{(-r)^{n-1}}{(n-1)!}
\Gamma(\nu n) \sin(\pi \nu n) \, ,
\ee
and it provides a generalization of the Gaussian and Airy functions:
\be
M_{1/2}(r)= \frac{1}{\sqrt{\pi}} \, \e^{-r^2/4} \, , 
\quad M_{1/3}(r)= 3^{2/3} Ai(r/3^{1/3}) \, .
\ee
Moreover, the following limit holds:
\be
\lim_{\nu \to 1^-} M_\nu(r) = \delta(r-1) \, .
\ee
The M density function is related to the Mittag--Leffler function through
the Laplace transform:
\be
\int_0^\infty \e^{-\lambda r} M_\nu(r) \, dr = E_\nu(-\lambda) \, ,
\ee
and it has an exponential decay for $r \to \infty$, i.e.:
\be
M_\nu(r) \sim \frac{Y^{\nu-1/2}}{\sqrt{2\pi}(1-\nu)^\nu \nu^{2\nu-1}}  
\, \e^{-Y} \,, \quad
Y=(1-\nu)(\nu^\nu r)^{1/(1-\nu)} \, ,
\ee
which allows for finite moments that can be computed through the formula:
\be
\int_0^\infty r^q M_\nu(r) dr = \frac{\Gamma(q+1)}{\Gamma(\nu q +1)} \,, \quad q > -1 \,.
\ee
A remarkable formula of the Mainardi density is the following integral representation
with $r \ge 0$, $0< \nu \,, \eta \,, \beta <1$ \cite{mainardi_etal-fcaa-2003}:
\be
M_\nu(r)=\int_0^\infty 
M_\eta\left(\frac{r}{\tau^\eta}\right) 
M_\beta(\tau) \, \frac{d\tau}{\tau^\eta} \, ; \quad \nu=\eta \beta \, ,
\ee
that, in the special case $\eta=1/2$, provides the following link with the Gaussian density:
\be
M_{\beta/2}(r)=\int_0^\infty 
\frac{\e^{-r^2/(4\tau)}}{\sqrt{\pi \tau}}   
M_\beta(\tau) \, \frac{d\tau}{\tau^\eta} \,.
\ee

\vspace{.2cm}
\noindent
The L\'evy stable density $L_\alpha^\theta(z)$, $-\infty < z < + \infty$,
$0 < \alpha < 2$, $|\theta|=\min\{\alpha, 2-\alpha\}$, is defined through the
Fourier transform:
\be
\int_{-\infty}^{+\infty} \e^{i \kappa z} L_\alpha^\theta(z) d\kappa =
\e^{-\Psi(\kappa)} \,, \quad \Psi(\kappa)=|\kappa|^\alpha \e^{i ({\rm{sgn}} \, \kappa)\theta \pi/2} \, .
\ee 
In the case $\theta=-\alpha$, $0 < \alpha < 1$, 
the L\'evy density reduces to a one-side density on the positive semi-axis
(when $\theta=\alpha$ on the negative semi-axis)
and it is defined through the Laplace transform:
\be
\int_0^\infty \e^{sz} L_\alpha^{-\alpha}(z) dz = \e^{-s^\alpha} \, .
\ee
The asymptotic behaviour for $|z| \to \infty$ is the power-law
\be
L_\alpha^\theta(z)=\mathcal{O}(|z|^{-(\alpha+1)} \, ,
\ee
and, for extremal densities, the following exponential decay holds
for $z \to 0$:
\be
L_\alpha^{-\alpha}(z) \sim
\frac{z^{-(2-\alpha)/(2(1-\alpha))}}{\sqrt{2\pi (1-\alpha)\alpha^{1/(\alpha-1)}}}
\, \e^{-Y} \,,
\quad Y=(1-\alpha)\alpha^{\alpha/(1-\alpha)} z^{\alpha/(1-\alpha)} \, .
\ee
Important special cases are the Gaussian, the Cauchy and the L\'evy--Smirnov density,
i.e.:
\be
L_2^0(z)= \frac{\e^{-z^2/4}}{2\sqrt{\pi}} \,, \quad
L_1^0(z)= \frac{1}{\pi}\frac{1}{1+z^2} \,, \quad
L_{1/2}^{-1/2}(z)= \frac{z^{-3/2}}{2\sqrt{\pi}} \, \e^{-1/(4z)}\, .
\ee
Moreover, the following limit holds:
\be
\lim_{\alpha \to 1} L_\alpha^{-\alpha}(z)=\delta(z-1) \, .
\ee
A remarkable formula of the L\'evy density is the following integral representation
for $z \ge 0$, $0< \beta < 1$:
\be
L_{\alpha_p}^{\theta_p}(z)=
\int_0^\infty L_{\alpha_q}^{\theta_q}\left(\frac{z}{\tau^{1/\theta_q}}\right) 
L_\beta^{-\beta}(\tau) 
\frac{d\tau}{\tau^{1/\alpha_q}} \,,
\quad \alpha_p=\beta \alpha_q \,, \quad \theta_p=\beta \theta_q \, ,
\ee
that, in the special case $\alpha_q=2$, $\theta_q=0$, 
provides the following link with the Gaussian density 
\cite{mainardi_etal-fcaa-2003,mainardi_etal-jms-2006}:
\be
L_{\alpha}^0(z)=
\int_0^\infty \frac{\e^{-z^2/(4\tau)}}{\sqrt{\pi \tau}} 
L_{\alpha/2}^{-\alpha/2}(\tau) d\tau \, . 
\ee 
The $M_\nu(r)$ function, $r \ge 0$, $0 < \nu < 1$,
and the extremal L\'evy density $L_\nu^{-\nu}(r)$ 
are related by the formula:
\be
\frac{1}{c^{1/\nu}} L_\nu^{-\nu}\left(\frac{r}{c^{1/\nu}}\right)=
\frac{c \, \nu}{r^{\nu+1}} M_\nu\left(\frac{c}{r^\nu}\right) \,, \quad
c > 0 \, .
\ee
In the present paper we consider such special densities in order to highlight
the relation of the proposed formulation with the fractional diffusion.
However, the asympototic behaviour of the modeled diffusion
can be achieved by using the asymptotic behaviour of the involved densities.
This means, by using exponential and power-law functions rather than special functions.

\section{Space-Time Fractional Diffusion}
\label{sec_stfde}

\noindent
For the particular choice of parameters: $\phi = 2\beta/\alpha\ ;\quad 1 < 
\phi < 2$, Eq. (\ref{random_nu}) reduces to the fundamental solution of the 
following Space-Time Fractional Diffusion equation:
\be 
{_tD_*^\beta} \, p(x;t) = 
A_{\alpha}\ 
{_xD_0^\alpha} \, p(x;t) \,, \quad -\infty < x < + \infty
\, , \quad t \ge 0 \, ,
\label{STFDE1}
\ee
with:
\be
A_{\alpha} = \left( C {\overline \nu} \right)^{\alpha/2}\, .
\label{gen_diff}
\ee
The nonlocal operators ${_tD_*^\beta}$ and ${_xD_0^\alpha}$ are the Caputo 
fractional time derivative and the Riesz-Feller space derivative, respectively
(see \cite{mainardi_etal-fcaa-2001} for the definition of these operators).
This is the same equation discussed in Refs. 
\cite{mainardi_etal-fcaa-2001,pagnini_etal-fcaa-2016}, but with a generalized 
fractional diffusivity $A_{\alpha}$ different from $1$.

\noindent
The solution reads:
\be
K_{\alpha,\beta}^{\theta}(x,t) = \frac{1}{ \left(A_{\alpha}\right)^{1/\alpha}
t^{\beta/\alpha}} K_{\alpha,\beta}^{\theta}
\left( \frac{x}{ \left(A_{\alpha}\right)^{1/\alpha}
  t^{\beta/\alpha}} \right)\, ,
\nonumber
\ee
with $\theta=0$ in this case.
\footnote{
Due to the self-similar property, here and in the following we use the same 
symbol for the two-variable function $F(x,t)$ and the associated one-variable
function written in terms of the similarity variable. Then, given the scaling
exponent $\Lambda$ and the coefficient $A$, 
we write: $F(x,t) = 1/t^\Lambda F(x/(A t^\Lambda))$.
This notation is not ambiguous as the meaning clearly follows from the number 
of independent variables.
}.
The superdiffusive regime determines the following constrain
on $\alpha$ and $\beta$: $\alpha/2 < \beta < \alpha$.
%

Given the solutions of the Time Fractional Diffusion equation
and of the Space Fractional Diffusion equation with diffusivity $1$
and $A_{\alpha}$, respectively
\cite{mainardi_etal-fcaa-2001,gorenflo_etal-cp-2002}:\\
$M_{\beta}(x,t)=1/t^\beta M_{\beta}(x/t^\beta)$ (Mainardi probability density)
and \\
$\Lat(x,t)=1/\left( A_{\alpha} t \right)^{1/\alpha} \Lat\left( x/
\left( A_{\alpha} t \right)^{1/\alpha} \right)$ (L\'evy probability density),\\
the general solution 
$K_{\alpha,\beta}^{\theta}$ can be written as a combination of these same solutions:
\be 
K_{\alpha,\beta}^{\theta}(x,t) = \int_0^\infty \Lat(x,\tau)
M_{\beta}\left(\tau,t\right) \, d\tau\, ,
\label{lumapa1}
\ee 
%
then the general solution emerges
as a linear combination of the temporal (Mainardi) and spatial (L\'evy)
solutions.
The Mainardi density is related to the extremal L\'evy density by the 
following relationship (see Section \ref{mainardi} for details):
\be 
\frac{t}{\beta \tau} \frac{1}{\tau^{1/\beta}} L_{\beta}^{-\beta}\left(\frac{t}{\tau^{1/\beta}}\right)
 = \frac{1}{t^\beta} \,
M_\beta\left(\frac{\tau}{t^\beta}\right) \,, \quad 0 < \beta \le 1 \,,
\quad \tau, t \ge 0 \,,
\label{LMformula}
\ee

\section*{Supplementary Material: References}


\end{document}